\documentclass[aps,prl,twocolumn,groupedaddress,letterpaper]{revtex4-1}

\usepackage{color}
\usepackage{graphicx}
\usepackage{float}
\usepackage{amsmath}
\usepackage[nolist,nohyperlinks]{acronym}
\usepackage{multirow}
\usepackage{lineno}
\usepackage{svn-multi}
\usepackage{import}
\usepackage[colorlinks=true,citecolor=red,urlcolor=red]{hyperref}
\usepackage[normalem]{ulem}
\usepackage[caption=false]{subfig}
\usepackage{pifont}
\usepackage{etoolbox}
\usepackage{xspace}
\usepackage[title, page]{appendix}
\usepackage{textcomp}

\newcommand{\pit}{\ensuremath{\pi^\mathrm{T}}}
\newcommand{\lalinference}{\textsc{LALInference}\xspace}

\newcommand{\mypiMaxPthirteen}{\ensuremath{3.113}}
\newcommand{\mypiThirteen}{\mypiMaxPthirteen\ensuremath{^{+0.049}_{-0.091}}\xspace}
\newcommand{\mypiBFthirteen}{\ensuremath{301}\xspace}

\newcommand{\mypiMaxPtwentyTwo}{\ensuremath{3.115}}
\newcommand{\mypiTwentyTwo}{\mypiMaxPtwentyTwo\ensuremath{^{+0.048}_{-0.088}}\xspace}
\newcommand{\mypiBFtwentyTwo}{\ensuremath{321}\xspace}

\begin{document}

\title{Pi from the sky -- A null test of general relativity from a population of gravitational wave observations}

\newcommand{\LIGOlabMIT}{\affiliation{LIGO Laboratory, Massachusetts Institute of Technology, 185 Albany St, Cambridge, MA 02139, USA}}
\newcommand{\MKI}{\affiliation{Department of Physics and Kavli Institute for Astrophysics and Space Research, Massachusetts Institute of Technology, 77 Massachusetts Ave, Cambridge, MA 02139, USA}}

\author{Carl-Johan Haster}  
\email[]{haster@mit.edu}
\LIGOlabMIT 
\MKI


\begin{abstract}
Our understanding of observed Gravitational Waves (GWs) comes from matching data to known signal models describing General Relativity (GR).
These models, expressed in the post-Newtonian formalism, contain the mathematical constant $\pi$.
Allowing $\pi$ to vary thus enables a strong, universal and generalisable null test of GR.
From a population of 22 GW observations, we make an astrophysical measurement of $\pi=\mypiTwentyTwo$, and prefer GR as the correct theory of gravity with a Bayes factor of \mypiBFtwentyTwo.
We find the variable $\pi$ test robust against simulated beyond-GR effects.
\end{abstract}

\maketitle

\acrodef{BBH}[BBH]{binary black hole}
\acrodef{BNS}[BNS]{binary neutron star}
\acrodef{BH}[BH]{black hole}
\acrodef{NS}[NS]{neutron star}
\acrodef{LVC}[LVC]{LIGO/Virgo Collaboration}
\acrodef{CBC}[CBC]{compact binary coalescence}
\acrodef{GR}[GR]{General Relativity}
\acrodef{CO}[CO]{compact object}
\acrodef{GW}[GW]{gravitational wave}
\acrodef{PN}[PN]{post-Newtonian}
\acrodef{PSD}[PSD]{power spectral density}
\acrodef{PDF}[PDF]{probability density function}
\acrodef{EoS}[EoS]{equation of state}
\acrodef{SPA}[SPA]{stationary phase approximation}
\acrodef{TF2}[\textsc{TF2}]{\textsc{TaylorF2}}
\acrodef{O1}[O1]{first observation run}
\acrodef{O2}[O2]{second observation run}
\acrodef{O3}[O3]{third observation run}
\acrodef{CI}[CI]{credible interval}
\acrodef{BF}[BF]{Bayes factor}
\acrodef{KDE}[KDE]{kernel density estimator}

\section{Introduction}
\label{sec:intro}

Observations of \acp{GW} from \acp{CBC} by the LIGO~\cite{Harry:2010zz} and Virgo~\cite{TheVirgo:2014hva} detectors have brought tests of \ac{GR} in the strong-field regime to hitherto unachievable levels~\cite{TheLIGOScientific:2016src,Abbott:2018lct,LIGOScientific:2019fpa,Isi:2019asy}.
This is fundamentally dependent on the detailed knowledge about the structure of the \acp{GW} emitted from a binary of \acp{CO} (\acp{BH} or \acp{NS}) stemming from decades of analytical~\cite{Blanchet:1995ez, Blanchet:2004ek, Blanchet:2005tk, Blanchet:2013haa, Buonanno:1998gg, Buonanno:2000ef, Damour:2008qf, Damour:2009kr, Barausse:2009xi, Damour:2016bks} and numerical~\cite{Pretorius:2005gq, Campanelli:2005dd, Baker:2005vv, Boyle:2019kee} studies of \acp{GW} from binary systems of \acp{CO}.

So far, the majority of theories of gravity beyond \ac{GR} are unable to construct predictions for \acp{GW} emitted by coalescing binaries with generic \acp{CO} (but see~\cite{Yagi:2011xp,Berti:2013gfa,Berti:2015itd,Hirschmann:2017psw,Okounkova:2017yby,Witek:2018dmd,Okounkova:2018pql,Loutrel:2018ydv,Okounkova:2019dfo,Torsello:2019tgc,Okounkova:2019zjf,Okounkova:2020rqw,Julie:2020vov, Witek:2020uzz} for the status of current efforts), 
hence tests of \ac{GR} are generically formulated as consistency tests only, where the primary approach is to introduce \textit{ad-hoc} modifications of the \ac{GR} waveforms.
This can be constructed to test different regions and functional dependencies of the overall waveform~\cite{Agathos:2013upa, Li:2011cg, Li:2011vx, Cornish:2011ys, Meidam:2017dgf, Abbott:2018lct}, such as deviations from the analytical coefficients of the \ac{PN} expansion~\cite{Blanchet:1995ez, Blanchet:2004ek, Blanchet:2005tk, Blanchet:2013haa}, which has been successful when investigating constraints on each included \ac{PN}-order separately~\cite{TheLIGOScientific:2016src, TheLIGOScientific:2016pea, Abbott:2017vtc, Abbott:2018lct, LIGOScientific:2019fpa,Isi:2019asy}.
These constraints can later be mapped onto bounds on specific alternate theories of gravity~\cite{Chatziioannou:2012rf,Yunes:2013dva,Yunes:2016jcc,Nair:2019iur}, something which in turn highlights a potential flaw of this approach.
Since the \ac{PN}-coefficients themselves depend on the specific properties of the source's \acp{CO}, like their masses, it would be reasonable to also assume any deviations from the \ac{GR}-predicted values to also be source dependent. 
If a hypothetical theory modifies \ac{BH}-spin behaviour, but not any mass parameters, a general \ac{PN}-deviation would be different for two binaries with the same \ac{BH} spin magnitudes but different mass ratios.
This is not accounted for in most current analyses~\cite{TheLIGOScientific:2016src, TheLIGOScientific:2016pea, Abbott:2017vtc, Abbott:2018lct, LIGOScientific:2019fpa} (but see ~\cite{Isi:2019asy} for a more general approach) and could lead to misinterpreted inference if any deviation from \ac{GR} was observed~\cite{Zimmerman:2019wzo,Purrer:2019jcp}.
In addition, the strength of this class of tests is reduced when more than one \ac{PN}-term is simultaneously allowed to vary, where the addition of a large number of unconstraining degrees of freedom generates an overall null gain in information about any of the included terms (cf. Fig 7 of~\cite{TheLIGOScientific:2016src}).

In this letter we implement a null test of \ac{GR}, probing the validity of the current knowledge about \ac{GR}, and specifically its nonlinear behaviour originating from \ac{GW} tail effects~\cite{Thorne:1980ru, Blanchet:1987wq, Blanchet:1992br, Blanchet:1993ec, Tanaka:1993pu, Blanchet:1994ez, Blanchet:1997jj}, with the mathematical constant $\pi$ treated as a variable.
$\pi$ can here be considered as a universal parameter across all \ac{GW} observations of \acp{CBC}~\cite{Zimmerman:2019wzo}, and simultaneously tests 4 (out of the included 8) \ac{PN}-orders.
This enables an unprecedentedly powerful test, as it is both theory-agnostic and conceptually generalisable to probe a population of \acp{GW} through a quantity that is formally consistent across independent observations (while also being comparatively inexpensive computationally).
Throughout this letter we denote the true value of $\pi$ as $\pit=3.141592653\ldots$~\cite{OEISpi}, a number which has been independently evaluated through several methods~\cite{Ramaley1969,Bailey1997,Galperin2003,Dumoulin:2014,Arndt2001,yCruncher}.
We assume $G = c = 1$.

\section{Method}
\label{sec:method}

The \ac{GW} signal from a \ac{CBC} can be generally expressed in the form
\begin{equation}
\tilde{h}(\theta,f) = A(\theta,f)e^{i\Psi(\theta,f)}\;,
\label{eq:htilde}
\end{equation}
where $\tilde{h}(f)$ is the emitted \ac{GW} strain in the frequency domain, with amplitude $A(\theta,f)$ and phase $\Psi(\theta,f)$ being functions of the source parameters $\theta$, e.g. \ac{CO} masses $m_{1,2}$, spin vectors $\vec{S}_{1,2}$ and tidal deformabilities $\Lambda_{1,2}$ (we fix $\Lambda_{1,2} = 0$ for \acp{BH}).
When the two \acp{CO} are sufficiently separated, for an orbital velocity $u \ll 1$ with $u = \pit M f$ and $M$ being the binary's total mass, Eq.~\eqref{eq:htilde} can be described accurately through a \acl{PN} expansion in $u$.
Under the \ac{SPA}~\cite{Sathyaprakash:1991mt,Cutler:1994ys,Apostolatos:1994mx,Poisson:1995ef,Droz:1999qx,Buonanno:2009zt}, $\Psi(\theta,f)$ is given for the \ac{TF2} model~\cite{Sathyaprakash:1991mt,Damour:2000zb, Damour:2002kr, Arun:2004hn, Buonanno:2009zt} as
\allowdisplaybreaks
\begin{equation}
\begin{split}
\Psi_{\text{TF2}} &= \, 2 \pit f t_c - \varphi_c - \pit/4\\
                        & + \frac{3}{128 \, \eta}u^{-5/3} \sum_{i=0}^{7}
\left(\varphi_i + \log(u)\varphi^l_i\right) u^{i/3}\;,
\end{split}
\label{eq:TF2phase}
\end{equation}
where $t_c$ and $\varphi_c$ are the overall time and phase defined at coalescence and $\eta = m_1m_2/(m_1 + m_2)^2$ is the symmetric mass ratio.
The expansion coefficients $\varphi_i$ and $\varphi^l_i$ are then given as functions of $\theta$~\cite{Sathyaprakash:1991mt,Buonanno:2009zt, Blanchet:2013haa, Bohe:2013cla, Poisson:1997ha, Arun:2008kb, Mikoczi:2005dn, Khan:2015jqa}.
In this formalism, multiples of $\pi$ appear in $\varphi_3$, $\varphi_5$, $\varphi^l_5$, $\varphi_6$ and  $\varphi_7$.
To also capture the post-inspiral \ac{PN} description of a \ac{CBC} signal, we employ the \textsc{IMRPhenomPv2} model, where the analytical inspiral description from \ac{TF2} is smoothly extended with a phenomenological description of a \ac{BBH} merger-ringdown section~\cite{Husa:2015iqa,Khan:2015jqa} together with an effective-precession treatment~\cite{Hannam:2013oca}.
For the \ac{BNS} events, this is further extended with a description of \ac{NS} matter effects~\cite{Dietrich:2018uni,Dietrich:2017aum}.
Neither extension depends on the variable $\pi$, as the extensions are phenomenological rather than analytical in their nature.

We note that $\pi$ is included in the orbital velocity $u$, originating in a conversion from angular to linear orbital frequencies.
From Eq.~\eqref{eq:TF2phase}, $u$ is already strongly constrained at the leading-order phase term and can be taken as known to sufficiently high precision in \ac{GR}.
Since we are here interested in specifically probing the \acl{PN} formalism, expressed through the $\varphi_i$ and $\varphi^l_i$ coefficients, we fix $\pi = \pit$ in $u$ throughout. 
Similarly, the $\pi$ in $2 \pit f t_c$, originating from a Fourier transform of the time-domain \ac{GW} signal, and the factor $\pit/4$ appears in Eq.~\eqref{eq:TF2phase} out of convention.
As both describe an overall phase shift, perfectly degenerate with the variables $t_c$ and $\varphi_c$ respectively,  we fix those two $\pi = \pit$ in this analysis.

It is important to note that the appearance of $\pi$ in the \ac{PN}-coefficients follows purely from definitions of mathematics itself~\cite{Blanchet:1997jj,Foffa:2011ub}, e.g. through the use of known identities to evaluate integrals of a specific form (cf. Eq. 5.4 of~\cite{Blanchet:1997jj}), and does not depend on the specific assumptions of \ac{GR} as a theory of gravity.
This formally justifies treating $\pi$ in all $\varphi_i$ and $\varphi^l_i$ coefficients as fundamentally the same quantity, and also treating it as a universal parameter across multiple independent \ac{CBC} observations~\cite{Zimmerman:2019wzo}.

We also note that the \ac{PN}-orders where $\pi$ appears are primarily describing so called \ac{GW} tail effects~\cite{Thorne:1980ru, Blanchet:1987wq, Blanchet:1992br, Blanchet:1993ec, Tanaka:1993pu, Blanchet:1994ez, Blanchet:1997jj}, where the outgoing \acp{GW} backscatter off the (approximately) static spacetime of the \ac{CBC} source.
Tail effects are an inherently nonlinear behaviour present in \ac{GR}, hence the use of a variable $\pi$ can directly probe the validity of the nonlinear terms expressed through the \ac{PN}-representation of \ac{GR} itself.

While the analysis with a variable $\pi$ formally is an extension of \ac{GR}, we do not argue that results presented here are direct suggestions for alternative theories of gravity.
Instead, we interpret this study primarily as a strong null test, validating the current understanding of \ac{GR} through a multi-order probe of the \ac{PN}-formalism.

Finally, we acknowledge that the \ac{GW} detectors, as well as the data they record, are constructed and calibrated for $\pi = \pit$ only.

\subsection{Bayesian methods}
\label{sec:bayesian}

We explore the parameter space $\theta$ defined by the \ac{CBC} models using Bayes' theorem to infer the posterior \ac{PDF}:
\begin{equation}
\label{Eq:BayesTheorem}
p(\theta|d, H) = \frac{p(\theta|H)p(d|\theta,H)}{p(d|H)}\;,
\end{equation}
where $p(\theta|H)$ is the prior \ac{PDF} of $\theta$ given the model $H$, $p(d|\theta,H)$ is the likelihood of observing the data $d$ assuming $\theta$ and $p(d|H) = \int p(\theta|H)p(d|\theta,H)d\theta$ is the evidence for $H$.
We preform Bayesian inference using the \lalinference package~\cite{Veitch:2014wba,LALInference-code, lalsuite}, following the analysis configuration from~\cite{LIGOScientific:2018mvr} which includes a fixed noise \ac{PSD} (defined for the analysed data $d$ and generated as a median \ac{PSD} using \textsc{BayesWave}~\cite{Littenberg:2014oda,Cornish:2014kda,Chatziioannou:2019zvs,GWTC1_PE_release,GW190425_PE_release,GW190412_PE_release}) and marginalisation over uncertainties in the calibration of $d$~\cite{SplineCalMarg-T1400682,GWTC1_PE_release,GW190425_PE_release,GW190412_PE_release,Cahillane:2017vkb}.
All GW events are analysed using publicly available data~\cite{GWOSC_O1, GWOSC_O2, GWOSC_GW190425, GWOSC_GW190412, Vallisneri:2014vxa, Abbott:2019ebz}.

We assume prior choices consistent with those used in~\cite{Abbott:2018exr,LIGOScientific:2018mvr,Abbott:2020uma,LIGOScientific:2020stg}.
For the two \acp{BNS}, we perform only analyses with $|\vec{S}_{1,2}| \leq 0.05$, and parametrize the \ac{NS} tidal deformability following the \acl{EoS} independent relations from~\cite{Chatziioannou:2018vzf}.
We assume a prior distribution for $\pi$ that is uniform between $-20 \leq \pi \leq 20$.

As $\pi$ can be considered a formally universal parameter, as defined by~\cite{Zimmerman:2019wzo}, it is trivial to evaluate joint constraints on $\pi$ from a set of $N$ individual observations by multiplying the $1D$ likelihood distributions (marginalised over all other parameters), dividing by one instance of the common prior and normalising the resulting posterior \ac{PDF}.

Finally, as \ac{GR} is nested inside the model which allows for a variable $\pi$ it is possible to compute a \ac{BF} in favour of \ac{GR}, more directly where $\pi = \pit$, using the Savage-Dickey density ratio ~\cite{dickey1971, BF_SavageDickey} as
\begin{equation}
\label{Eq:SavageDickey}
\mathrm{BF} = \frac{p(\pi = \pit|d, H) }{p(\pi = \pit|H)}\; ,
\end{equation}
i.e. the ratio of the posterior and prior \acp{PDF} evaluated at $\pi = \pit$.

\section{Astrophysical measurement of $\pi$}
\label{sec:realpi}

\begin{figure}
\centering
\includegraphics[width=\columnwidth]{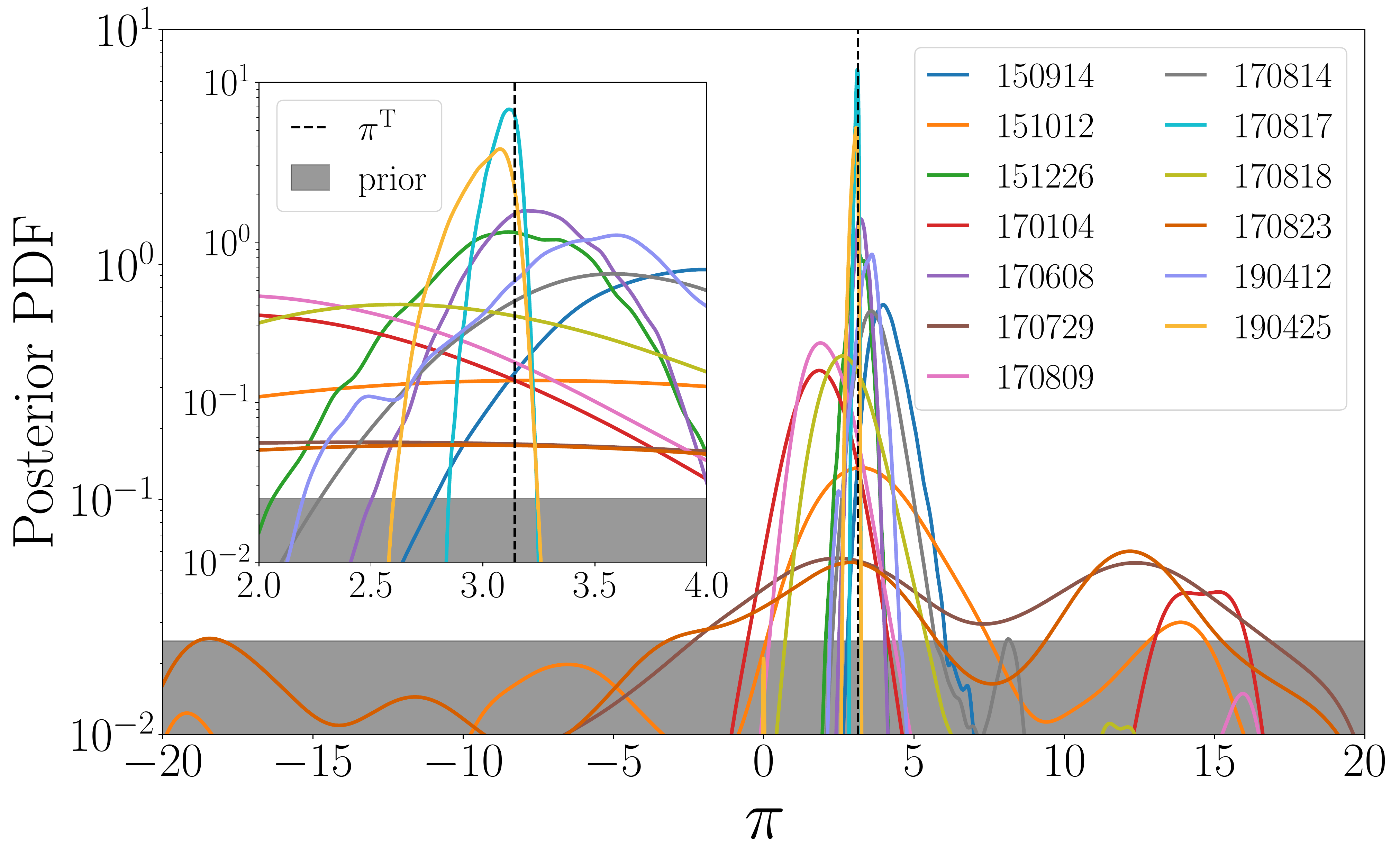}
\caption{Individual posterior \acp{PDF} of $\pi$ for the thirteen \ac{LVC} \ac{GW} observations~\cite{LIGOScientific:2018mvr, Abbott:2020uma,LIGOScientific:2020stg}.
The dashed line indicates $\pit$, the true value of $\pi$.
The shaded gray region indicates the prior \ac{PDF}.
}
\label{fig:IndPi}
\end{figure}

The \ac{LVC} has so far, from its \ac{O1}, \ac{O2} and \ac{O3}~\cite{Aasi:2013wya}, confirmed 13 \ac{GW} observations, 2 \acp{BNS} and 11 \acp{BBH}~\cite{LIGOScientific:2018mvr,Abbott:2020uma,LIGOScientific:2020stg}.
Whereas other studies restrict themselves to high-significance events only \cite{LIGOScientific:2019fpa,Isi:2019asy}, primarily due to computational restrictions, the analysis presented here is easily extendable to and informed by all available \ac{GW} observations.
The individual-event posterior \acp{PDF} for $\pi$ are shown in Fig.~\ref{fig:IndPi}, visualised through \acp{KDE}. 
All \ac{GW} events support the region near $\pit$, with the strongest constraints coming from the two \acp{BNS} and the lowest-mass \acp{BBH} (GW151226 and GW170806).
This agrees with prior expectations as lower-mass \ac{CBC} signals are dominated by the binary inspiral, described by the \ac{PN}-series, in turn constrained by this analysis.
Apart from a general broadening of the recovered posterior \acp{PDF} in other source parameters, consistent with the addition of a new degree of freedom, we note no general degeneracies between $\pi$ and other parameters.
This is especially noticeable as $\varphi_3$ contains the leading-order terms for both $\pi$ and the effects from \ac{CO}-spins.
As both $\pi$ and spin-parameters however appear jointly at higher \ac{PN}-orders, with different interdependences than in $\varphi_3$, the potentially strong degeneracy is thus broken in this analysis.

\begin{figure}
\centering
\includegraphics[width=\columnwidth]{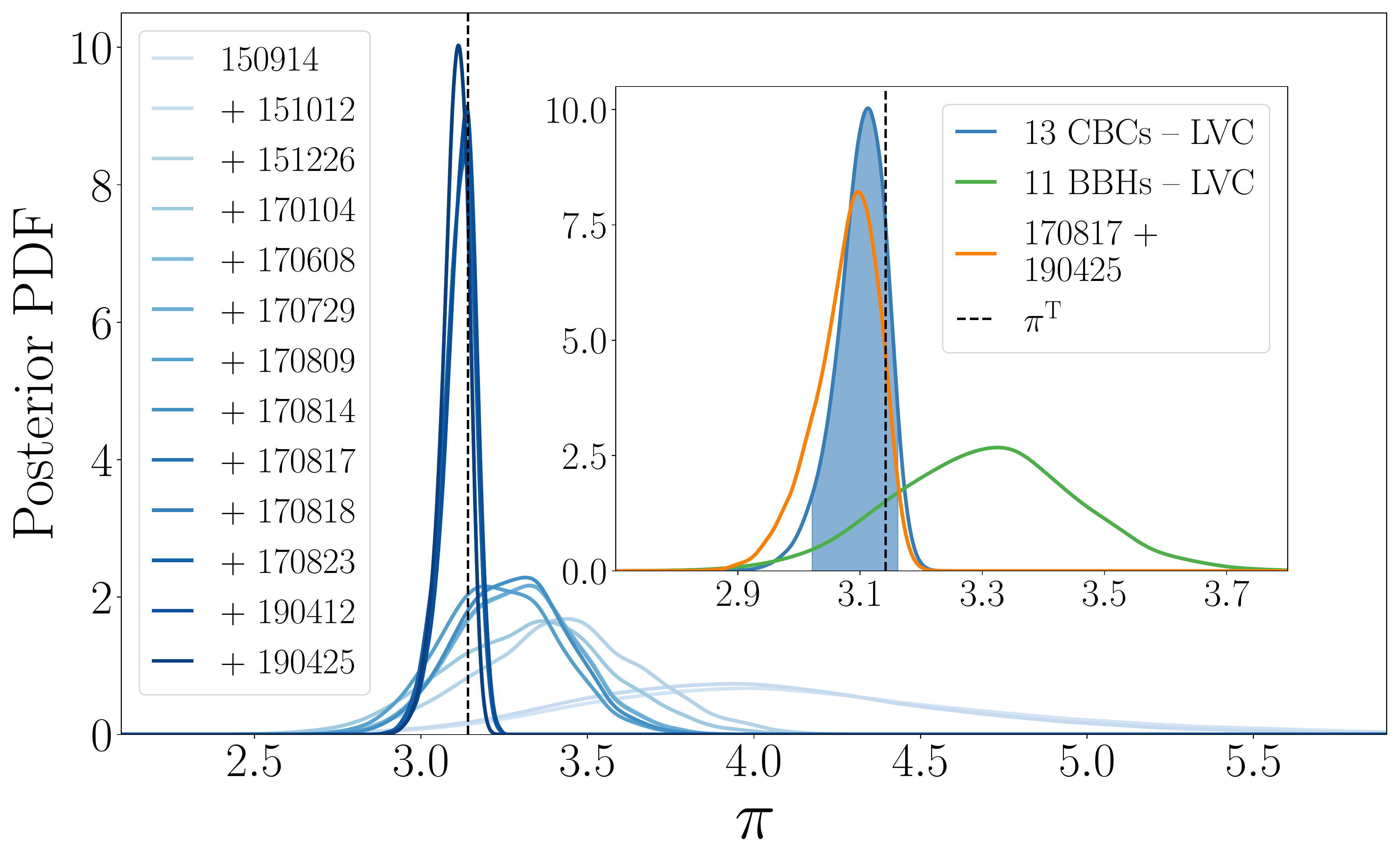}
\caption{Joint posterior \acp{PDF} of $\pi$ for the thirteen \ac{LVC} \ac{GW} observations~\cite{LIGOScientific:2018mvr, Abbott:2020uma,LIGOScientific:2020stg}.
The observations are added chronologically, corrresponding to the light to dark blue transition.
The dashed line indicates $\pit$, the true value of $\pi$.
\textit{Inset}: The joint posterior \acp{PDF} for the eleven \acp{BBH} (green), the two \acp{BNS} (orange) and all 13 \ac{LVC} \acp{CBC}.
The shaded blue region in the inset corresponds to the $90\%$ \ac{CI}.
}
\label{fig:JoinPi}
\end{figure}

The chronological progression of the joint posterior \ac{PDF} of $\pi$, from the population of \acp{CBC} reported by the \ac{LVC} is shown in Fig.~\ref{fig:JoinPi}, again highlighting the significant contribution of the four lowest-mass events.
Together, these 13 events give a maximum a posteriori value, with associated $90\%$ \ac{CI}, of $\pi=$\mypiThirteen.
For this set of events, the \acl{BF} in favour of \ac{GR} being an accurate description of strong-field gravity is \mypiBFthirteen.

\begin{figure}
\centering
\includegraphics[width=\columnwidth]{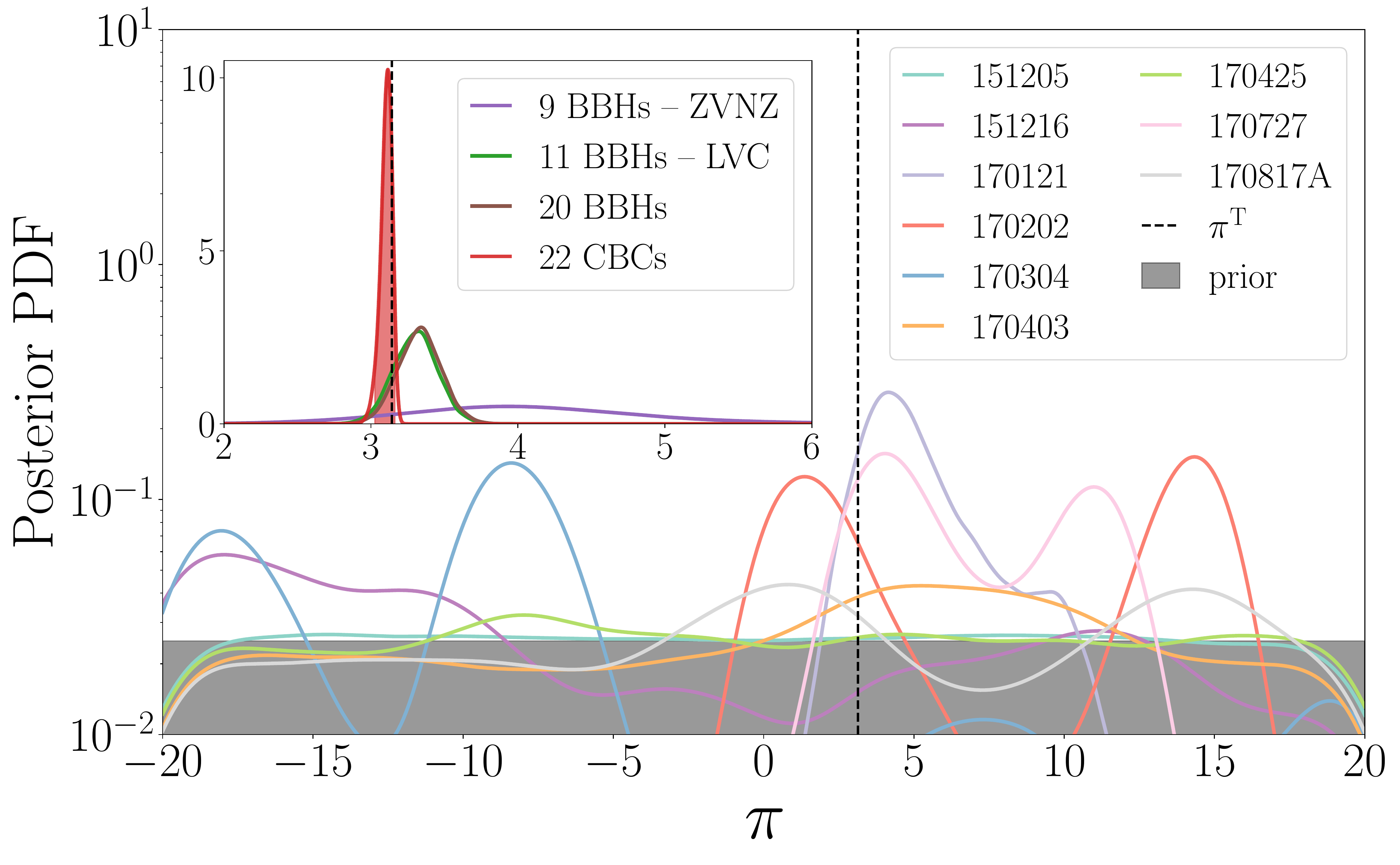}
\caption{Individual posterior \acp{PDF} of $\pi$ for the nine ZVNZ \acp{BBH} observations~\cite{Zackay:2019tzo,Venumadhav:2019lyq,Nitz:2019hdf,Zackay:2019btq}.
The dashed line indicates $\pit$, the true value of $\pi$.
The shaded gray region indicates the prior \ac{PDF}.
\textit{Inset}: The joint posterior \acp{PDF} for the nine ZVNZ \acp{BBH} (purple), the eleven \ac{LVC} \acp{BBH} (green, cf. Fig.~\ref{fig:JoinPi}), all 20 \acp{BBH} (brown) and all 22 \acp{CBC} (red).
The shaded red region in the inset corresponds to the $90\%$ \ac{CI}.
}
\label{fig:ExtraPi}
\end{figure}

In addition to the eleven \acp{BBH} reported by the \ac{LVC} in~\cite{LIGOScientific:2018mvr}, independent analyses (\citet{Zackay:2019tzo,Venumadhav:2019lyq,Nitz:2019hdf,Zackay:2019btq} hereafter collectively labelled ZVNZ) have claimed an additional nine \ac{BBH} observations, whose posterior \acp{PDF} of $\pi$ are shown in Fig.~\ref{fig:ExtraPi}.
It should be noted that out of all 22 included \acp{CBC}, only GW151216~\cite{Zackay:2019tzo} and GW170304~\cite{Venumadhav:2019lyq} recover $\pi$ disfavouring $\pit$ with single-event \acp{BF} in support of \ac{GR} of $1/2$ and $1/5$ respectively.
These two events have previously been identified as especially sensitive to overall prior and analysis choices~\cite{Galaudage:2019jdx,Huang:2020ysn}. 
The population of 22 \ac{CBC} observations gives a measurement of $\pi=$\mypiTwentyTwo, and a \ac{BF} in favour of \ac{GR} of \mypiBFtwentyTwo.
This constitutes the strongest constraints on the validity of the positive \ac{PN}-order coefficients to date~\cite{MinusTwoPN}, with a fractional width of the joint $\pi$ $90\%$ \ac{CI} $< 0.04$, more than a factor of 2 improvement over previous single-\ac{PN}-order variability results~\cite{Abbott:2018lct,LIGOScientific:2019fpa,GWTC-1_TGR_dataRelease}.
We also note a more significant improvement when comparing against the constraints on $\varphi_3$, the lowest \ac{PN}-order directly probed by this analysis.
This can be attributed to a combination of the inherent multi-order nature of the variable $\pi$ analysis and the inclusion of a larger population of \ac{CBC} observations than previous studies, thus together enabling a stronger constraint on the validity of the tested theory.

\section{\ac{BBH}-like noise transients -- $\pi$ estimation}
\label{sec:background}

\begin{figure}
\centering
\includegraphics[width=\columnwidth]{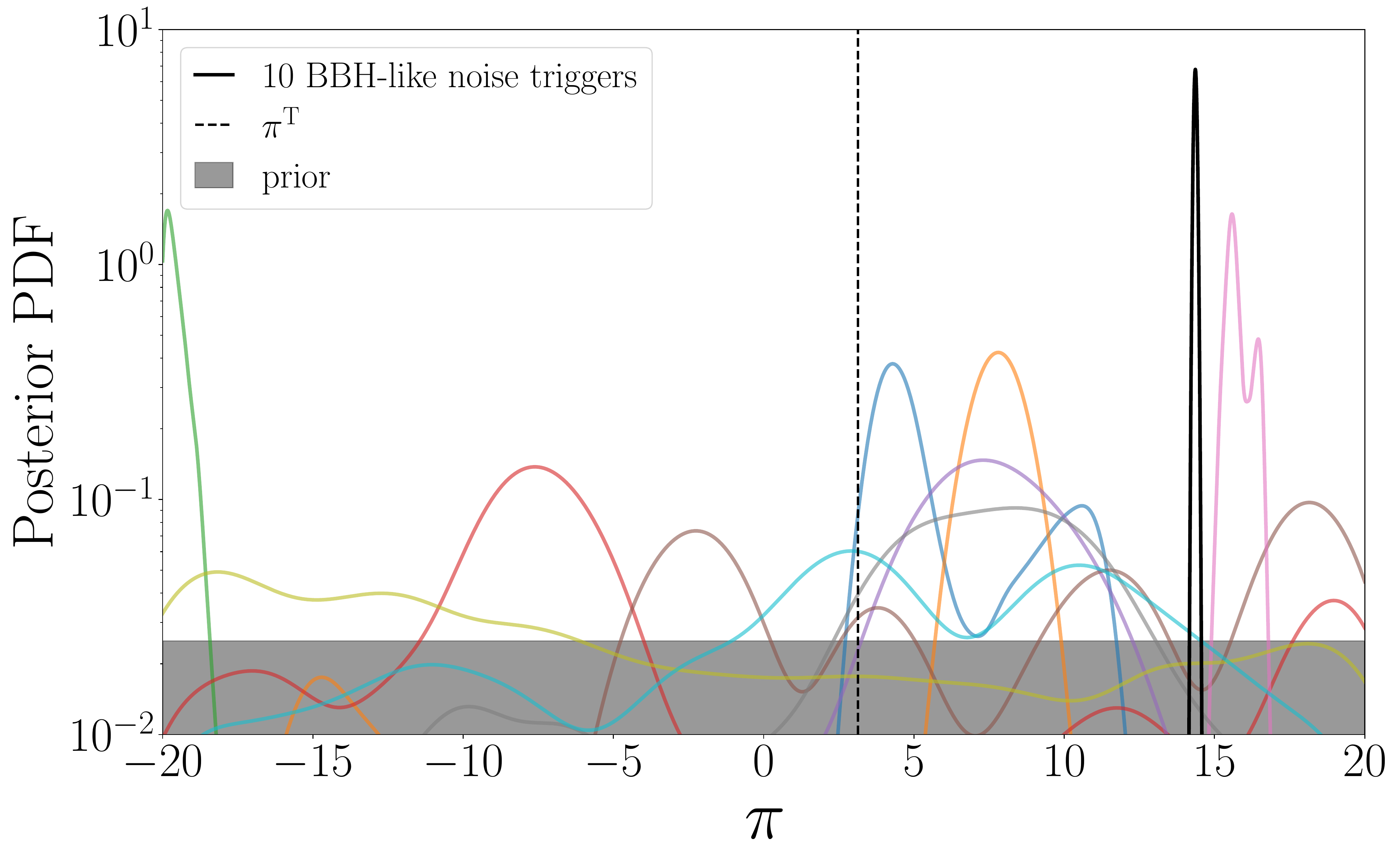}
\caption{Individual posterior \acp{PDF} (coloured) of $\pi$ for ten \ac{BBH}-like noise triggers from a background distribution.
The dashed line indicates $\pit$, the true value of $\pi$.
The shaded gray region indicates the prior\ac{PDF}.
The black line corresponds to the joint posterior \acp{PDF} for the background triggers.
}
\label{fig:background}
\end{figure}

In order to test the reliability of this analysis against spurious false-positives we analyse a set of background triggers, where sections of real data from LIGO~\cite{GWOSC_O1, Abbott:2019ebz} have been offset in time by longer than the light-travel time between sites.
This time-shifted data is thus guaranteed to not contain any real coincident \ac{GW} events, and primarily represent noise-transients from the LIGO instruments.
We select 10 high-significance BBH background triggers produced by the \textsc{PyCBC} search pipeline~\cite{Nitz:2017svb, Usman:2015kfa, Canton:2014ena, PyCBC_zenodo} and used by~\cite{Isi:2018vst}.
Following the same procedure as above, we recover their individual and joint posterior \acp{PDF} on $\pi$, shown in Fig.~\ref{fig:background}.
When the analysed signals do not correspond to \ac{GR}, as is the case for these noise-transients, there is significant scatter of the recovered posterior \acp{PDF} in $\pi$.
The apparent strong constraint from the joint analysis of the 10 triggers stems primarily from the narrow region hosting the black curve in Fig.~\ref{fig:background} being the only range where $p(\pi|d, H) \neq 0$ for all triggers, and that the joint posterior \ac{PDF} has a unit area.
The corresponding \ac{BF} $\lesssim 1/10^{300}$ for $\pi = \pit$, further highlighting the variable $\pi$ analysis correctly identifying the non-\ac{GR} features present in these noise triggers.
We conclude that the analysis appears stable against generating false-positive results when exposed to even a small population of 10 known non-\ac{GR} (background noise) signals.

\section{Presence of massive graviton}
\label{sec:lambdag}

To show that this analysis can reveal the presence of realistic beyond-\ac{GR} effects, we simulate three \ac{BBH} systems with parameters consistent with the three \acp{BBH} detected by the \ac{LVC} during \ac{O1}~\cite{TheLIGOScientific:2016pea, LIGOScientific:2018mvr, GWTC1_PE_release}.
We modify the $\varphi_2$ \ac{PN} coefficients of the simulated signals to mimic a massive graviton with a Compton wavelength $\lambda_G$~\cite{Will:1997bb,Zimmerman:2019wzo}.
We choose $\lambda_G$ to be in the range between $2.48\times10^{13}\:\mathrm{km}$, consistent with the current \ac{GW} observational lower bound from~\cite{LIGOScientific:2019fpa}, and $10^{12}\:\mathrm{km}$, a value already ruled-out observationally.
We also simulate ``pure'' \ac{GR}, with $\lambda_G = \infty$.
It should be noted that $\lambda_G$ enters at a \ac{PN}-order where $\pi$ is not present.

\begin{figure}
\centering
\includegraphics[width=\columnwidth]{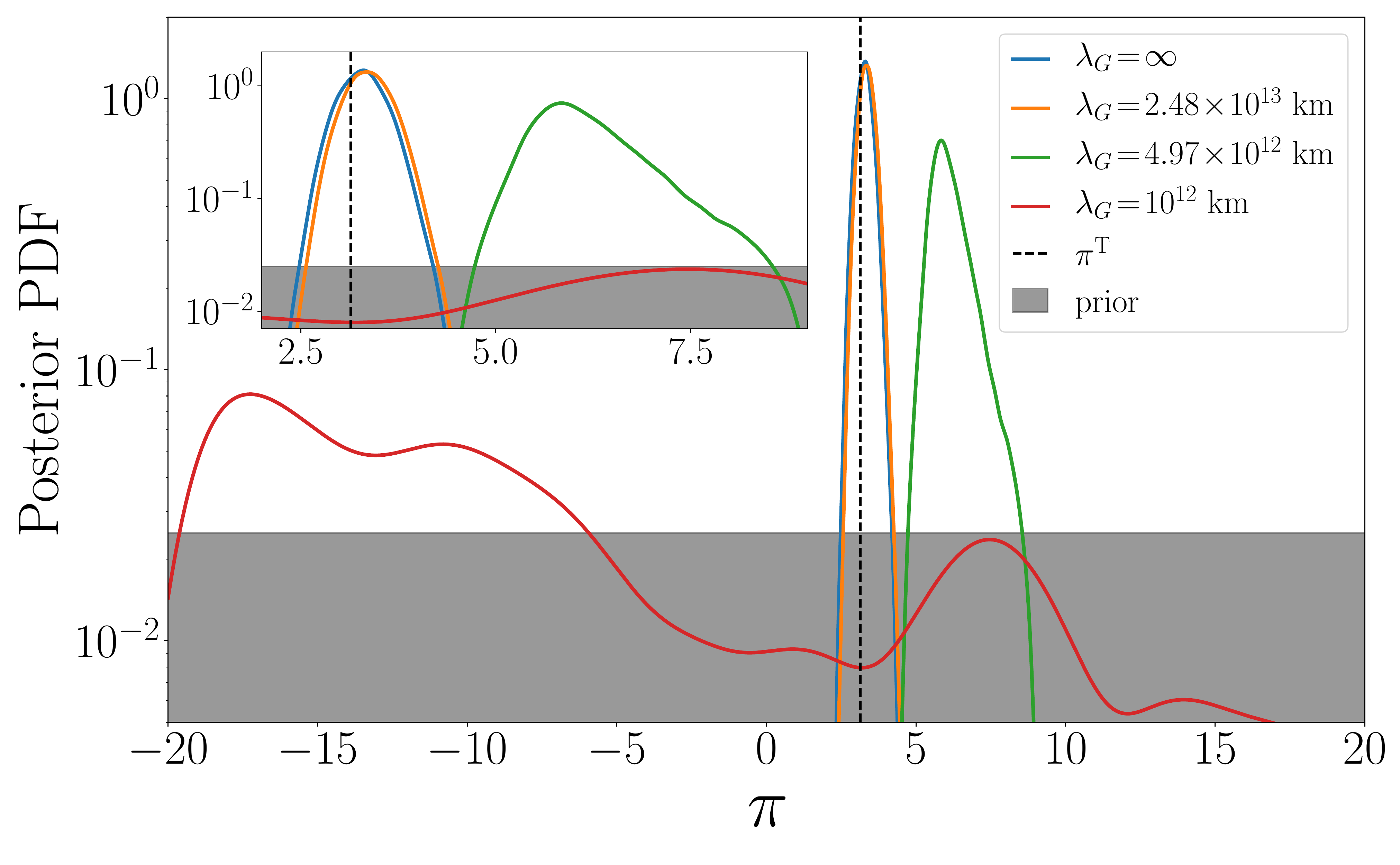}
\caption{Joint distributions of $\pi$ shown for four different instances of $\lambda_G$, for \ac{GW} signals consistent with the three \ac{O1} events}
\label{fig:lambdag}
\end{figure}

In Fig.~\ref{fig:lambdag} we show the joint posterior \acp{PDF} on $\pi$ from the three \ac{BBH} signals for each value of $\lambda_G$.
Given that $\lambda_G \geq 2.48\times10^{13}\:\mathrm{km}$ is not ruled out by current \ac{GW} observations~\cite{LIGOScientific:2019fpa}, it is not surprising that an analysis using this bound yields a posterior \ac{PDF} in agreement with \ac{GR}.
For $\lambda_G = 10^{12}\:\mathrm{km}$, more than an order of magnitude below the current lower bound, the recovered $\pi$ posterior \ac{PDF} is biased away from $\pit$ but only marginally informative over the assumed prior. 
This indicates that a strong beyond-\ac{GR} effect, acting partially orthogonal to the changes to the signal from a varying $\pi$, can be sufficient to saturate the constraining power of this test.
Namely, if no allowed value of $\pi$ is able to sufficiently ``correct'' for the beyond-\ac{GR} modification present in the signal, the variable $\pi$ degree of freedom becomes uninformative.
It is instead the case in between these extremes that is the most illustrative, where a presence of a marginal beyond-\ac{GR} effect induces a clear bias in the recovered $\pi$ and a \ac{BF}$\sim 1/10^{15}$ for $\pi = \pit$.
Hence, a detection of $\pi \neq \pit$ in a population of real observations can be interpreted as first indication of the presence of beyond-\ac{GR} behaviour, with the variable $\pi$ test being especially powerful from its generalisable and multi-\ac{PN}-order nature.
The identification of $\pi \neq \pit$ does itself not guide what beyond-\ac{GR} effect is present.
Such questions can only be answered by performing theory-specific model comparison analyses~\cite{Yunes:2013dva,Yunes:2016jcc, Abbott:2018lct,LIGOScientific:2019fpa} over the population of observations for which $\pi \neq \pit$.

\section{Discussion}
\label{sec:conclusions}

In the \acl{PN} formalism of \acl{GR}, the mathematical constant $\pi$ presents a powerful null test of our currently preferred theory of gravity.
With $\pi$ simultaneously probing four \ac{PN}-orders, fundamentally describing the same conceptual quantity in all instances, doing so in a theory-agnostic way that is also generalisable and universal across independent \ac{GW} observations, it provides an unmatched capability for validating our understanding of \ac{GR}. 
Using the current set of 22 \ac{CBC} observations in data from LIGO and Virgo, identified by both the \ac{LVC} and independent researchers~\cite{LIGOScientific:2018mvr,Abbott:2020uma,LIGOScientific:2020stg,Zackay:2019tzo,Venumadhav:2019lyq,Nitz:2019hdf,Zackay:2019btq}, we achieve an astrophysical measurement of $\pi=$\mypiTwentyTwo, consistent with the accepted $\pit$ value.
This is the most stringent constraint on the positive-order \ac{PN}-series to date~\cite{MinusTwoPN}, and the first viable multi-\ac{PN}-order constraint from \ac{GW} observations.
The analysis also allows, through the construction of a \acl{BF}, direct validation of \ac{GR} with \ac{BF}=\mypiBFtwentyTwo in support of it as the currently favoured theory of gravity.
We have shown the analysis to be robust when exposed to non-signal, but high-significance, \ac{GW} triggers as well as being able to indicate the presence of beyond-\ac{GR} effects in the case where such signals were to exist.
The method presented in this letter is easily extended to future \ac{GW} observations, of both \acp{CBC} and other modelled sources, such as the quasi-monochromatic \acp{GW} emitted by spinning \acp{NS}~\cite{Pisarski:2019vxw, Abbott:2019bed, Authors:2019ztc}, and capable of accommodating observations from across the \ac{GW} spectrum~\cite{Berti:2015itd, Sesana:2016ljz, Vitale:2016rfr, Chamberlain:2017fjl, Carson:2019kkh, Carson:2019rda, Marsat:2020rtl, Toubiana:2020vtf,Kelley:2017vox, Burke-Spolaor:2018bvk, Aggarwal:2018mgp,Becsy:2019dim,Gnocchi:2019jzp}.

\begin{acknowledgments}
The author thanks Katerina Chatziioannou, Riccardo Sturani, Salvatore Vitale and Aaron Zimmerman for helpful suggestions and discussion.
I also thank Maximiliano Isi, and the authors of~\cite{Isi:2018vst}, for providing access to the background BBH-like trigger information.
The author acknowledges support of the National Science Foundation, and the LIGO Laboratory. 
LIGO was constructed by the California Institute of Technology and Massachusetts Institute of Technology with funding from the National Science Foundation and operates under cooperative agreement PHY-1764464. 
The author for computational resources provided by the LIGO Laboratory and supported by National Science Foundation Grants PHY-0757058 and PHY-0823459.
This research has made use of data, software and/or web tools obtained from the Gravitational Wave Open Science Center (\url{https://www.gw-openscience.org}), a service of LIGO Laboratory, the LIGO Scientific Collaboration and the Virgo Collaboration. 
LIGO is funded by the U.S. National Science Foundation. 
Virgo is funded by the French Centre National de Recherche Scientifique (CNRS), the Italian Istituto Nazionale della Fisica Nucleare (INFN) and the Dutch Nikhef, with contributions by Polish and Hungarian institutes.
This analysis was made possible by the {\tt LALSuite}~\cite{lalsuite}, {\tt numpy}~\cite{numpy}, {\tt SciPy}~\cite{Virtanen:2019joe} and {\tt matplotlib}~\cite{Hunter:2007ouj} software packages. 
This is LIGO Document Number DCC-P2000159.
\end{acknowledgments}

\bibliography{PiFromTheSky.bib}

\begin{thebibliography}{137}%
\makeatletter
\providecommand \@ifxundefined [1]{%
 \@ifx{#1\undefined}
}%
\providecommand \@ifnum [1]{%
 \ifnum #1\expandafter \@firstoftwo
 \else \expandafter \@secondoftwo
 \fi
}%
\providecommand \@ifx [1]{%
 \ifx #1\expandafter \@firstoftwo
 \else \expandafter \@secondoftwo
 \fi
}%
\providecommand \natexlab [1]{#1}%
\providecommand \enquote  [1]{``#1''}%
\providecommand \bibnamefont  [1]{#1}%
\providecommand \bibfnamefont [1]{#1}%
\providecommand \citenamefont [1]{#1}%
\providecommand \href@noop [0]{\@secondoftwo}%
\providecommand \href [0]{\begingroup \@sanitize@url \@href}%
\providecommand \@href[1]{\@@startlink{#1}\@@href}%
\providecommand \@@href[1]{\endgroup#1\@@endlink}%
\providecommand \@sanitize@url [0]{\catcode `\\12\catcode `\$12\catcode
  `\&12\catcode `\#12\catcode `\^12\catcode `\_12\catcode `\%12\relax}%
\providecommand \@@startlink[1]{}%
\providecommand \@@endlink[0]{}%
\providecommand \url  [0]{\begingroup\@sanitize@url \@url }%
\providecommand \@url [1]{\endgroup\@href {#1}{\urlprefix }}%
\providecommand \urlprefix  [0]{URL }%
\providecommand \Eprint [0]{\href }%
\providecommand \doibase [0]{http://dx.doi.org/}%
\providecommand \selectlanguage [0]{\@gobble}%
\providecommand \bibinfo  [0]{\@secondoftwo}%
\providecommand \bibfield  [0]{\@secondoftwo}%
\providecommand \translation [1]{[#1]}%
\providecommand \BibitemOpen [0]{}%
\providecommand \bibitemStop [0]{}%
\providecommand \bibitemNoStop [0]{.\EOS\space}%
\providecommand \EOS [0]{\spacefactor3000\relax}%
\providecommand \BibitemShut  [1]{\csname bibitem#1\endcsname}%
\let\auto@bib@innerbib\@empty
\bibitem [{\citenamefont {Harry}(2010)}]{Harry:2010zz}%
  \BibitemOpen
  \bibfield  {author} {\bibinfo {author} {\bibfnamefont {G.~M.}\ \bibnamefont
  {Harry}} (\bibinfo {collaboration} {LIGO Scientific Collaboration}),\ }\href
  {\doibase 10.1088/0264-9381/27/8/084006} {\bibfield  {journal} {\bibinfo
  {journal} {Class.\ Quant.\ Grav.}\ }\textbf {\bibinfo {volume} {27}},\
  \bibinfo {pages} {084006} (\bibinfo {year} {2010})}\BibitemShut {NoStop}%
\bibitem [{\citenamefont {Acernese}\ \emph {et~al.}(2015)\citenamefont
  {Acernese} \emph {et~al.}}]{TheVirgo:2014hva}%
  \BibitemOpen
  \bibfield  {author} {\bibinfo {author} {\bibfnamefont {F.}~\bibnamefont
  {Acernese}} \emph {et~al.} (\bibinfo {collaboration} {Virgo Collaboration}),\
  }\href {\doibase 10.1088/0264-9381/32/2/024001} {\bibfield  {journal}
  {\bibinfo  {journal} {Class.\ Quant.\ Grav.}\ }\textbf {\bibinfo {volume}
  {32}},\ \bibinfo {pages} {024001} (\bibinfo {year} {2015})},\ \Eprint
  {http://arxiv.org/abs/1408.3978} {arXiv:1408.3978 [gr-qc]} \BibitemShut
  {NoStop}%
\bibitem [{\citenamefont {Abbott}\ \emph
  {et~al.}(2016{\natexlab{a}})\citenamefont {Abbott} \emph
  {et~al.}}]{TheLIGOScientific:2016src}%
  \BibitemOpen
  \bibfield  {author} {\bibinfo {author} {\bibfnamefont {B.}~\bibnamefont
  {Abbott}} \emph {et~al.} (\bibinfo {collaboration} {LIGO Scientific
  Collaboration, Virgo Collaboration}),\ }\href {\doibase
  10.1103/PhysRevLett.116.221101} {\bibfield  {journal} {\bibinfo  {journal}
  {Phys.\ Rev.\ Lett.}\ }\textbf {\bibinfo {volume} {116}},\ \bibinfo {pages}
  {221101} (\bibinfo {year} {2016}{\natexlab{a}})},\ \bibinfo {note} {[Erratum:
  Phys.Rev.Lett. 121, 129902 (2018)]},\ \Eprint
  {http://arxiv.org/abs/1602.03841} {arXiv:1602.03841 [gr-qc]} \BibitemShut
  {NoStop}%
\bibitem [{\citenamefont {Abbott}\ \emph
  {et~al.}(2019{\natexlab{a}})\citenamefont {Abbott} \emph
  {et~al.}}]{Abbott:2018lct}%
  \BibitemOpen
  \bibfield  {author} {\bibinfo {author} {\bibfnamefont {B.}~\bibnamefont
  {Abbott}} \emph {et~al.} (\bibinfo {collaboration} {LIGO Scientific
  Collaboration, Virgo Collaboration}),\ }\href {\doibase
  10.1103/PhysRevLett.123.011102} {\bibfield  {journal} {\bibinfo  {journal}
  {Phys.\ Rev.\ Lett.}\ }\textbf {\bibinfo {volume} {123}},\ \bibinfo {pages}
  {011102} (\bibinfo {year} {2019}{\natexlab{a}})},\ \Eprint
  {http://arxiv.org/abs/1811.00364} {arXiv:1811.00364 [gr-qc]} \BibitemShut
  {NoStop}%
\bibitem [{\citenamefont {Abbott}\ \emph
  {et~al.}(2019{\natexlab{b}})\citenamefont {Abbott} \emph
  {et~al.}}]{LIGOScientific:2019fpa}%
  \BibitemOpen
  \bibfield  {author} {\bibinfo {author} {\bibfnamefont {B.}~\bibnamefont
  {Abbott}} \emph {et~al.} (\bibinfo {collaboration} {LIGO Scientific
  Collaboration, Virgo Collaboration}),\ }\href {\doibase
  10.1103/PhysRevD.100.104036} {\bibfield  {journal} {\bibinfo  {journal}
  {Phys.\ Rev.\ D}\ }\textbf {\bibinfo {volume} {100}},\ \bibinfo {pages}
  {104036} (\bibinfo {year} {2019}{\natexlab{b}})},\ \Eprint
  {http://arxiv.org/abs/1903.04467} {arXiv:1903.04467 [gr-qc]} \BibitemShut
  {NoStop}%
\bibitem [{\citenamefont {Isi}\ \emph {et~al.}(2019)\citenamefont {Isi},
  \citenamefont {Chatziioannou},\ and\ \citenamefont {Farr}}]{Isi:2019asy}%
  \BibitemOpen
  \bibfield  {author} {\bibinfo {author} {\bibfnamefont {M.}~\bibnamefont
  {Isi}}, \bibinfo {author} {\bibfnamefont {K.}~\bibnamefont {Chatziioannou}},
  \ and\ \bibinfo {author} {\bibfnamefont {W.~M.}\ \bibnamefont {Farr}},\
  }\href {\doibase 10.1103/PhysRevLett.123.121101} {\bibfield  {journal}
  {\bibinfo  {journal} {Phys. Rev. Lett.}\ }\textbf {\bibinfo {volume} {123}},\
  \bibinfo {pages} {121101} (\bibinfo {year} {2019})},\ \Eprint
  {http://arxiv.org/abs/1904.08011} {arXiv:1904.08011 [gr-qc]} \BibitemShut
  {NoStop}%
\bibitem [{\citenamefont {Blanchet}\ \emph {et~al.}(1995)\citenamefont
  {Blanchet}, \citenamefont {Damour}, \citenamefont {Iyer}, \citenamefont
  {Will},\ and\ \citenamefont {Wiseman}}]{Blanchet:1995ez}%
  \BibitemOpen
  \bibfield  {author} {\bibinfo {author} {\bibfnamefont {L.}~\bibnamefont
  {Blanchet}}, \bibinfo {author} {\bibfnamefont {T.}~\bibnamefont {Damour}},
  \bibinfo {author} {\bibfnamefont {B.~R.}\ \bibnamefont {Iyer}}, \bibinfo
  {author} {\bibfnamefont {C.~M.}\ \bibnamefont {Will}}, \ and\ \bibinfo
  {author} {\bibfnamefont {A.~G.}\ \bibnamefont {Wiseman}},\ }\href {\doibase
  10.1103/PhysRevLett.74.3515} {\bibfield  {journal} {\bibinfo  {journal}
  {Phys.\ Rev.\ Lett.}\ }\textbf {\bibinfo {volume} {74}},\ \bibinfo {pages}
  {3515} (\bibinfo {year} {1995})},\ \Eprint
  {http://arxiv.org/abs/gr-qc/9501027} {arXiv:gr-qc/9501027} \BibitemShut
  {NoStop}%
\bibitem [{\citenamefont {Blanchet}\ \emph {et~al.}(2004)\citenamefont
  {Blanchet}, \citenamefont {Damour}, \citenamefont {Esposito-Farese},\ and\
  \citenamefont {Iyer}}]{Blanchet:2004ek}%
  \BibitemOpen
  \bibfield  {author} {\bibinfo {author} {\bibfnamefont {L.}~\bibnamefont
  {Blanchet}}, \bibinfo {author} {\bibfnamefont {T.}~\bibnamefont {Damour}},
  \bibinfo {author} {\bibfnamefont {G.}~\bibnamefont {Esposito-Farese}}, \ and\
  \bibinfo {author} {\bibfnamefont {B.~R.}\ \bibnamefont {Iyer}},\ }\href
  {\doibase 10.1103/PhysRevLett.93.091101} {\bibfield  {journal} {\bibinfo
  {journal} {Phys.\ Rev.\ Lett.}\ }\textbf {\bibinfo {volume} {93}},\ \bibinfo
  {pages} {091101} (\bibinfo {year} {2004})},\ \Eprint
  {http://arxiv.org/abs/gr-qc/0406012} {arXiv:gr-qc/0406012} \BibitemShut
  {NoStop}%
\bibitem [{\citenamefont {Blanchet}\ \emph {et~al.}(2005)\citenamefont
  {Blanchet}, \citenamefont {Damour}, \citenamefont {Esposito-Farese},\ and\
  \citenamefont {Iyer}}]{Blanchet:2005tk}%
  \BibitemOpen
  \bibfield  {author} {\bibinfo {author} {\bibfnamefont {L.}~\bibnamefont
  {Blanchet}}, \bibinfo {author} {\bibfnamefont {T.}~\bibnamefont {Damour}},
  \bibinfo {author} {\bibfnamefont {G.}~\bibnamefont {Esposito-Farese}}, \ and\
  \bibinfo {author} {\bibfnamefont {B.~R.}\ \bibnamefont {Iyer}},\ }\href
  {\doibase 10.1103/PhysRevD.71.124004} {\bibfield  {journal} {\bibinfo
  {journal} {Phys.\ Rev.\ D}\ }\textbf {\bibinfo {volume} {71}},\ \bibinfo
  {pages} {124004} (\bibinfo {year} {2005})},\ \Eprint
  {http://arxiv.org/abs/gr-qc/0503044} {arXiv:gr-qc/0503044} \BibitemShut
  {NoStop}%
\bibitem [{\citenamefont {Blanchet}(2014)}]{Blanchet:2013haa}%
  \BibitemOpen
  \bibfield  {author} {\bibinfo {author} {\bibfnamefont {L.}~\bibnamefont
  {Blanchet}},\ }\href {\doibase 10.12942/lrr-2014-2} {\bibfield  {journal}
  {\bibinfo  {journal} {Living Rev.\ Rel.}\ }\textbf {\bibinfo {volume} {17}},\
  \bibinfo {pages} {2} (\bibinfo {year} {2014})},\ \Eprint
  {http://arxiv.org/abs/1310.1528} {arXiv:1310.1528 [gr-qc]} \BibitemShut
  {NoStop}%
\bibitem [{\citenamefont {Buonanno}\ and\ \citenamefont
  {Damour}(1999)}]{Buonanno:1998gg}%
  \BibitemOpen
  \bibfield  {author} {\bibinfo {author} {\bibfnamefont {A.}~\bibnamefont
  {Buonanno}}\ and\ \bibinfo {author} {\bibfnamefont {T.}~\bibnamefont
  {Damour}},\ }\href {\doibase 10.1103/PhysRevD.59.084006} {\bibfield
  {journal} {\bibinfo  {journal} {Phys.\ Rev.\ D}\ }\textbf {\bibinfo {volume}
  {59}},\ \bibinfo {pages} {084006} (\bibinfo {year} {1999})},\ \Eprint
  {http://arxiv.org/abs/gr-qc/9811091} {arXiv:gr-qc/9811091} \BibitemShut
  {NoStop}%
\bibitem [{\citenamefont {Buonanno}\ and\ \citenamefont
  {Damour}(2000)}]{Buonanno:2000ef}%
  \BibitemOpen
  \bibfield  {author} {\bibinfo {author} {\bibfnamefont {A.}~\bibnamefont
  {Buonanno}}\ and\ \bibinfo {author} {\bibfnamefont {T.}~\bibnamefont
  {Damour}},\ }\href {\doibase 10.1103/PhysRevD.62.064015} {\bibfield
  {journal} {\bibinfo  {journal} {Phys.\ Rev.\ D}\ }\textbf {\bibinfo {volume}
  {62}},\ \bibinfo {pages} {064015} (\bibinfo {year} {2000})},\ \Eprint
  {http://arxiv.org/abs/gr-qc/0001013} {arXiv:gr-qc/0001013} \BibitemShut
  {NoStop}%
\bibitem [{\citenamefont {Damour}\ \emph {et~al.}(2008)\citenamefont {Damour},
  \citenamefont {Jaranowski},\ and\ \citenamefont {Schaefer}}]{Damour:2008qf}%
  \BibitemOpen
  \bibfield  {author} {\bibinfo {author} {\bibfnamefont {T.}~\bibnamefont
  {Damour}}, \bibinfo {author} {\bibfnamefont {P.}~\bibnamefont {Jaranowski}},
  \ and\ \bibinfo {author} {\bibfnamefont {G.}~\bibnamefont {Schaefer}},\
  }\href {\doibase 10.1103/PhysRevD.78.024009} {\bibfield  {journal} {\bibinfo
  {journal} {Phys.\ Rev.\ D}\ }\textbf {\bibinfo {volume} {78}},\ \bibinfo
  {pages} {024009} (\bibinfo {year} {2008})},\ \Eprint
  {http://arxiv.org/abs/0803.0915} {arXiv:0803.0915 [gr-qc]} \BibitemShut
  {NoStop}%
\bibitem [{\citenamefont {Damour}\ and\ \citenamefont
  {Nagar}(2009)}]{Damour:2009kr}%
  \BibitemOpen
  \bibfield  {author} {\bibinfo {author} {\bibfnamefont {T.}~\bibnamefont
  {Damour}}\ and\ \bibinfo {author} {\bibfnamefont {A.}~\bibnamefont {Nagar}},\
  }\href {\doibase 10.1103/PhysRevD.79.081503} {\bibfield  {journal} {\bibinfo
  {journal} {Phys.\ Rev.\ D}\ }\textbf {\bibinfo {volume} {79}},\ \bibinfo
  {pages} {081503} (\bibinfo {year} {2009})},\ \Eprint
  {http://arxiv.org/abs/0902.0136} {arXiv:0902.0136 [gr-qc]} \BibitemShut
  {NoStop}%
\bibitem [{\citenamefont {Barausse}\ and\ \citenamefont
  {Buonanno}(2010)}]{Barausse:2009xi}%
  \BibitemOpen
  \bibfield  {author} {\bibinfo {author} {\bibfnamefont {E.}~\bibnamefont
  {Barausse}}\ and\ \bibinfo {author} {\bibfnamefont {A.}~\bibnamefont
  {Buonanno}},\ }\href {\doibase 10.1103/PhysRevD.81.084024} {\bibfield
  {journal} {\bibinfo  {journal} {Phys.\ Rev.\ D}\ }\textbf {\bibinfo {volume}
  {81}},\ \bibinfo {pages} {084024} (\bibinfo {year} {2010})},\ \Eprint
  {http://arxiv.org/abs/0912.3517} {arXiv:0912.3517 [gr-qc]} \BibitemShut
  {NoStop}%
\bibitem [{\citenamefont {Damour}\ and\ \citenamefont
  {Nagar}(2016)}]{Damour:2016bks}%
  \BibitemOpen
  \bibfield  {author} {\bibinfo {author} {\bibfnamefont {T.}~\bibnamefont
  {Damour}}\ and\ \bibinfo {author} {\bibfnamefont {A.}~\bibnamefont {Nagar}},\
  }\enquote {\bibinfo {title} {The effective-one-body approach to the general
  relativistic two body problem},}\ in\ \href {\doibase
  10.1007/978-3-319-19416-5_7} {\emph {\bibinfo {booktitle} {Astrophysical
  Black Holes, Lecture Notes in Physics}}},\ Vol.\ \bibinfo {volume} {905}\
  (\bibinfo  {publisher} {Springer International Publishing},\ \bibinfo {year}
  {2016})\ pp.\ \bibinfo {pages} {273--312}\BibitemShut {NoStop}%
\bibitem [{\citenamefont {Pretorius}(2005)}]{Pretorius:2005gq}%
  \BibitemOpen
  \bibfield  {author} {\bibinfo {author} {\bibfnamefont {F.}~\bibnamefont
  {Pretorius}},\ }\href {\doibase 10.1103/PhysRevLett.95.121101} {\bibfield
  {journal} {\bibinfo  {journal} {Phys.\ Rev.\ Lett.}\ }\textbf {\bibinfo
  {volume} {95}},\ \bibinfo {pages} {121101} (\bibinfo {year} {2005})},\
  \Eprint {http://arxiv.org/abs/gr-qc/0507014} {arXiv:gr-qc/0507014}
  \BibitemShut {NoStop}%
\bibitem [{\citenamefont {Campanelli}\ \emph {et~al.}(2006)\citenamefont
  {Campanelli}, \citenamefont {Lousto}, \citenamefont {Marronetti},\ and\
  \citenamefont {Zlochower}}]{Campanelli:2005dd}%
  \BibitemOpen
  \bibfield  {author} {\bibinfo {author} {\bibfnamefont {M.}~\bibnamefont
  {Campanelli}}, \bibinfo {author} {\bibfnamefont {C.}~\bibnamefont {Lousto}},
  \bibinfo {author} {\bibfnamefont {P.}~\bibnamefont {Marronetti}}, \ and\
  \bibinfo {author} {\bibfnamefont {Y.}~\bibnamefont {Zlochower}},\ }\href
  {\doibase 10.1103/PhysRevLett.96.111101} {\bibfield  {journal} {\bibinfo
  {journal} {Phys.\ Rev.\ Lett.}\ }\textbf {\bibinfo {volume} {96}},\ \bibinfo
  {pages} {111101} (\bibinfo {year} {2006})},\ \Eprint
  {http://arxiv.org/abs/gr-qc/0511048} {arXiv:gr-qc/0511048} \BibitemShut
  {NoStop}%
\bibitem [{\citenamefont {Baker}\ \emph {et~al.}(2006)\citenamefont {Baker},
  \citenamefont {Centrella}, \citenamefont {Choi}, \citenamefont {Koppitz},\
  and\ \citenamefont {van Meter}}]{Baker:2005vv}%
  \BibitemOpen
  \bibfield  {author} {\bibinfo {author} {\bibfnamefont {J.~G.}\ \bibnamefont
  {Baker}}, \bibinfo {author} {\bibfnamefont {J.}~\bibnamefont {Centrella}},
  \bibinfo {author} {\bibfnamefont {D.-I.}\ \bibnamefont {Choi}}, \bibinfo
  {author} {\bibfnamefont {M.}~\bibnamefont {Koppitz}}, \ and\ \bibinfo
  {author} {\bibfnamefont {J.}~\bibnamefont {van Meter}},\ }\href {\doibase
  10.1103/PhysRevLett.96.111102} {\bibfield  {journal} {\bibinfo  {journal}
  {Phys.\ Rev.\ Lett.}\ }\textbf {\bibinfo {volume} {96}},\ \bibinfo {pages}
  {111102} (\bibinfo {year} {2006})},\ \Eprint
  {http://arxiv.org/abs/gr-qc/0511103} {arXiv:gr-qc/0511103} \BibitemShut
  {NoStop}%
\bibitem [{\citenamefont {Boyle}\ \emph {et~al.}(2019)\citenamefont {Boyle}
  \emph {et~al.}}]{Boyle:2019kee}%
  \BibitemOpen
  \bibfield  {author} {\bibinfo {author} {\bibfnamefont {M.}~\bibnamefont
  {Boyle}} \emph {et~al.},\ }\href {\doibase 10.1088/1361-6382/ab34e2}
  {\bibfield  {journal} {\bibinfo  {journal} {Class.\ Quant.\ Grav.}\ }\textbf
  {\bibinfo {volume} {36}},\ \bibinfo {pages} {195006} (\bibinfo {year}
  {2019})},\ \Eprint {http://arxiv.org/abs/1904.04831} {arXiv:1904.04831
  [gr-qc]} \BibitemShut {NoStop}%
\bibitem [{\citenamefont {Yagi}\ \emph {et~al.}(2012)\citenamefont {Yagi},
  \citenamefont {Stein}, \citenamefont {Yunes},\ and\ \citenamefont
  {Tanaka}}]{Yagi:2011xp}%
  \BibitemOpen
  \bibfield  {author} {\bibinfo {author} {\bibfnamefont {K.}~\bibnamefont
  {Yagi}}, \bibinfo {author} {\bibfnamefont {L.~C.}\ \bibnamefont {Stein}},
  \bibinfo {author} {\bibfnamefont {N.}~\bibnamefont {Yunes}}, \ and\ \bibinfo
  {author} {\bibfnamefont {T.}~\bibnamefont {Tanaka}},\ }\href {\doibase
  10.1103/PhysRevD.85.064022} {\bibfield  {journal} {\bibinfo  {journal} {Phys.
  Rev. D}\ }\textbf {\bibinfo {volume} {85}},\ \bibinfo {pages} {064022}
  (\bibinfo {year} {2012})},\ \bibinfo {note} {[Erratum: Phys.Rev.D 93, 029902
  (2016)]},\ \Eprint {http://arxiv.org/abs/1110.5950} {arXiv:1110.5950 [gr-qc]}
  \BibitemShut {NoStop}%
\bibitem [{\citenamefont {Berti}\ \emph {et~al.}(2013)\citenamefont {Berti},
  \citenamefont {Cardoso}, \citenamefont {Gualtieri}, \citenamefont
  {Horbatsch},\ and\ \citenamefont {Sperhake}}]{Berti:2013gfa}%
  \BibitemOpen
  \bibfield  {author} {\bibinfo {author} {\bibfnamefont {E.}~\bibnamefont
  {Berti}}, \bibinfo {author} {\bibfnamefont {V.}~\bibnamefont {Cardoso}},
  \bibinfo {author} {\bibfnamefont {L.}~\bibnamefont {Gualtieri}}, \bibinfo
  {author} {\bibfnamefont {M.}~\bibnamefont {Horbatsch}}, \ and\ \bibinfo
  {author} {\bibfnamefont {U.}~\bibnamefont {Sperhake}},\ }\href {\doibase
  10.1103/PhysRevD.87.124020} {\bibfield  {journal} {\bibinfo  {journal} {Phys.
  Rev. D}\ }\textbf {\bibinfo {volume} {87}},\ \bibinfo {pages} {124020}
  (\bibinfo {year} {2013})},\ \Eprint {http://arxiv.org/abs/1304.2836}
  {arXiv:1304.2836 [gr-qc]} \BibitemShut {NoStop}%
\bibitem [{\citenamefont {Berti}\ \emph {et~al.}(2015)\citenamefont {Berti}
  \emph {et~al.}}]{Berti:2015itd}%
  \BibitemOpen
  \bibfield  {author} {\bibinfo {author} {\bibfnamefont {E.}~\bibnamefont
  {Berti}} \emph {et~al.},\ }\href {\doibase 10.1088/0264-9381/32/24/243001}
  {\bibfield  {journal} {\bibinfo  {journal} {Class. Quant. Grav.}\ }\textbf
  {\bibinfo {volume} {32}},\ \bibinfo {pages} {243001} (\bibinfo {year}
  {2015})},\ \Eprint {http://arxiv.org/abs/1501.07274} {arXiv:1501.07274
  [gr-qc]} \BibitemShut {NoStop}%
\bibitem [{\citenamefont {Hirschmann}\ \emph {et~al.}(2018)\citenamefont
  {Hirschmann}, \citenamefont {Lehner}, \citenamefont {Liebling},\ and\
  \citenamefont {Palenzuela}}]{Hirschmann:2017psw}%
  \BibitemOpen
  \bibfield  {author} {\bibinfo {author} {\bibfnamefont {E.~W.}\ \bibnamefont
  {Hirschmann}}, \bibinfo {author} {\bibfnamefont {L.}~\bibnamefont {Lehner}},
  \bibinfo {author} {\bibfnamefont {S.~L.}\ \bibnamefont {Liebling}}, \ and\
  \bibinfo {author} {\bibfnamefont {C.}~\bibnamefont {Palenzuela}},\ }\href
  {\doibase 10.1103/PhysRevD.97.064032} {\bibfield  {journal} {\bibinfo
  {journal} {Phys. Rev. D}\ }\textbf {\bibinfo {volume} {97}},\ \bibinfo
  {pages} {064032} (\bibinfo {year} {2018})},\ \Eprint
  {http://arxiv.org/abs/1706.09875} {arXiv:1706.09875 [gr-qc]} \BibitemShut
  {NoStop}%
\bibitem [{\citenamefont {Okounkova}\ \emph {et~al.}(2017)\citenamefont
  {Okounkova}, \citenamefont {Stein}, \citenamefont {Scheel},\ and\
  \citenamefont {Hemberger}}]{Okounkova:2017yby}%
  \BibitemOpen
  \bibfield  {author} {\bibinfo {author} {\bibfnamefont {M.}~\bibnamefont
  {Okounkova}}, \bibinfo {author} {\bibfnamefont {L.~C.}\ \bibnamefont
  {Stein}}, \bibinfo {author} {\bibfnamefont {M.~A.}\ \bibnamefont {Scheel}}, \
  and\ \bibinfo {author} {\bibfnamefont {D.~A.}\ \bibnamefont {Hemberger}},\
  }\href {\doibase 10.1103/PhysRevD.96.044020} {\bibfield  {journal} {\bibinfo
  {journal} {Phys. Rev. D}\ }\textbf {\bibinfo {volume} {96}},\ \bibinfo
  {pages} {044020} (\bibinfo {year} {2017})},\ \Eprint
  {http://arxiv.org/abs/1705.07924} {arXiv:1705.07924 [gr-qc]} \BibitemShut
  {NoStop}%
\bibitem [{\citenamefont {Witek}\ \emph {et~al.}(2019)\citenamefont {Witek},
  \citenamefont {Gualtieri}, \citenamefont {Pani},\ and\ \citenamefont
  {Sotiriou}}]{Witek:2018dmd}%
  \BibitemOpen
  \bibfield  {author} {\bibinfo {author} {\bibfnamefont {H.}~\bibnamefont
  {Witek}}, \bibinfo {author} {\bibfnamefont {L.}~\bibnamefont {Gualtieri}},
  \bibinfo {author} {\bibfnamefont {P.}~\bibnamefont {Pani}}, \ and\ \bibinfo
  {author} {\bibfnamefont {T.~P.}\ \bibnamefont {Sotiriou}},\ }\href {\doibase
  10.1103/PhysRevD.99.064035} {\bibfield  {journal} {\bibinfo  {journal} {Phys.
  Rev. D}\ }\textbf {\bibinfo {volume} {99}},\ \bibinfo {pages} {064035}
  (\bibinfo {year} {2019})},\ \Eprint {http://arxiv.org/abs/1810.05177}
  {arXiv:1810.05177 [gr-qc]} \BibitemShut {NoStop}%
\bibitem [{\citenamefont {Okounkova}\ \emph
  {et~al.}(2019{\natexlab{a}})\citenamefont {Okounkova}, \citenamefont
  {Scheel},\ and\ \citenamefont {Teukolsky}}]{Okounkova:2018pql}%
  \BibitemOpen
  \bibfield  {author} {\bibinfo {author} {\bibfnamefont {M.}~\bibnamefont
  {Okounkova}}, \bibinfo {author} {\bibfnamefont {M.~A.}\ \bibnamefont
  {Scheel}}, \ and\ \bibinfo {author} {\bibfnamefont {S.~A.}\ \bibnamefont
  {Teukolsky}},\ }\href {\doibase 10.1103/PhysRevD.99.044019} {\bibfield
  {journal} {\bibinfo  {journal} {Phys. Rev. D}\ }\textbf {\bibinfo {volume}
  {99}},\ \bibinfo {pages} {044019} (\bibinfo {year} {2019}{\natexlab{a}})},\
  \Eprint {http://arxiv.org/abs/1811.10713} {arXiv:1811.10713 [gr-qc]}
  \BibitemShut {NoStop}%
\bibitem [{\citenamefont {Loutrel}\ \emph {et~al.}(2018)\citenamefont
  {Loutrel}, \citenamefont {Tanaka},\ and\ \citenamefont
  {Yunes}}]{Loutrel:2018ydv}%
  \BibitemOpen
  \bibfield  {author} {\bibinfo {author} {\bibfnamefont {N.}~\bibnamefont
  {Loutrel}}, \bibinfo {author} {\bibfnamefont {T.}~\bibnamefont {Tanaka}}, \
  and\ \bibinfo {author} {\bibfnamefont {N.}~\bibnamefont {Yunes}},\ }\href
  {\doibase 10.1103/PhysRevD.98.064020} {\bibfield  {journal} {\bibinfo
  {journal} {Phys. Rev. D}\ }\textbf {\bibinfo {volume} {98}},\ \bibinfo
  {pages} {064020} (\bibinfo {year} {2018})},\ \Eprint
  {http://arxiv.org/abs/1806.07431} {arXiv:1806.07431 [gr-qc]} \BibitemShut
  {NoStop}%
\bibitem [{\citenamefont {Okounkova}\ \emph
  {et~al.}(2019{\natexlab{b}})\citenamefont {Okounkova}, \citenamefont {Stein},
  \citenamefont {Scheel},\ and\ \citenamefont {Teukolsky}}]{Okounkova:2019dfo}%
  \BibitemOpen
  \bibfield  {author} {\bibinfo {author} {\bibfnamefont {M.}~\bibnamefont
  {Okounkova}}, \bibinfo {author} {\bibfnamefont {L.~C.}\ \bibnamefont
  {Stein}}, \bibinfo {author} {\bibfnamefont {M.~A.}\ \bibnamefont {Scheel}}, \
  and\ \bibinfo {author} {\bibfnamefont {S.~A.}\ \bibnamefont {Teukolsky}},\
  }\href {\doibase 10.1103/PhysRevD.100.104026} {\bibfield  {journal} {\bibinfo
   {journal} {Phys. Rev. D}\ }\textbf {\bibinfo {volume} {100}},\ \bibinfo
  {pages} {104026} (\bibinfo {year} {2019}{\natexlab{b}})},\ \Eprint
  {http://arxiv.org/abs/1906.08789} {arXiv:1906.08789 [gr-qc]} \BibitemShut
  {NoStop}%
\bibitem [{\citenamefont {Torsello}\ \emph {et~al.}(2020)\citenamefont
  {Torsello}, \citenamefont {Kocic}, \citenamefont {H{\"o}g{\r{a}}s},\ and\
  \citenamefont {M{\"o}rtsell}}]{Torsello:2019tgc}%
  \BibitemOpen
  \bibfield  {author} {\bibinfo {author} {\bibfnamefont {F.}~\bibnamefont
  {Torsello}}, \bibinfo {author} {\bibfnamefont {M.}~\bibnamefont {Kocic}},
  \bibinfo {author} {\bibfnamefont {M.}~\bibnamefont {H{\"o}g{\r{a}}s}}, \ and\
  \bibinfo {author} {\bibfnamefont {E.}~\bibnamefont {M{\"o}rtsell}},\ }\href
  {\doibase 10.1088/1361-6382/ab56fc} {\bibfield  {journal} {\bibinfo
  {journal} {Class. Quant. Grav.}\ }\textbf {\bibinfo {volume} {37}},\ \bibinfo
  {pages} {025013} (\bibinfo {year} {2020})},\ \Eprint
  {http://arxiv.org/abs/1904.07869} {arXiv:1904.07869 [gr-qc]} \BibitemShut
  {NoStop}%
\bibitem [{\citenamefont {Okounkova}\ \emph {et~al.}(2020)\citenamefont
  {Okounkova}, \citenamefont {Stein}, \citenamefont {Moxon}, \citenamefont
  {Scheel},\ and\ \citenamefont {Teukolsky}}]{Okounkova:2019zjf}%
  \BibitemOpen
  \bibfield  {author} {\bibinfo {author} {\bibfnamefont {M.}~\bibnamefont
  {Okounkova}}, \bibinfo {author} {\bibfnamefont {L.~C.}\ \bibnamefont
  {Stein}}, \bibinfo {author} {\bibfnamefont {J.}~\bibnamefont {Moxon}},
  \bibinfo {author} {\bibfnamefont {M.~A.}\ \bibnamefont {Scheel}}, \ and\
  \bibinfo {author} {\bibfnamefont {S.~A.}\ \bibnamefont {Teukolsky}},\ }\href
  {\doibase 10.1103/PhysRevD.101.104016} {\bibfield  {journal} {\bibinfo
  {journal} {Phys. Rev. D}\ }\textbf {\bibinfo {volume} {101}},\ \bibinfo
  {pages} {104016} (\bibinfo {year} {2020})},\ \Eprint
  {http://arxiv.org/abs/1911.02588} {arXiv:1911.02588 [gr-qc]} \BibitemShut
  {NoStop}%
\bibitem [{\citenamefont {Okounkova}(2020)}]{Okounkova:2020rqw}%
  \BibitemOpen
  \bibfield  {author} {\bibinfo {author} {\bibfnamefont {M.}~\bibnamefont
  {Okounkova}},\ }\href@noop {} {\  (\bibinfo {year} {2020})},\ \Eprint
  {http://arxiv.org/abs/2001.03571} {arXiv:2001.03571 [gr-qc]} \BibitemShut
  {NoStop}%
\bibitem [{\citenamefont {Juli{\'e}}\ and\ \citenamefont
  {Berti}(2020)}]{Julie:2020vov}%
  \BibitemOpen
  \bibfield  {author} {\bibinfo {author} {\bibfnamefont {F.-L.}\ \bibnamefont
  {Juli{\'e}}}\ and\ \bibinfo {author} {\bibfnamefont {E.}~\bibnamefont
  {Berti}},\ }\href@noop {} {\  (\bibinfo {year} {2020})},\ \Eprint
  {http://arxiv.org/abs/2004.00003} {arXiv:2004.00003 [gr-qc]} \BibitemShut
  {NoStop}%
\bibitem [{\citenamefont {Witek}\ \emph {et~al.}(2020)\citenamefont {Witek},
  \citenamefont {Gualtieri},\ and\ \citenamefont {Pani}}]{Witek:2020uzz}%
  \BibitemOpen
  \bibfield  {author} {\bibinfo {author} {\bibfnamefont {H.}~\bibnamefont
  {Witek}}, \bibinfo {author} {\bibfnamefont {L.}~\bibnamefont {Gualtieri}}, \
  and\ \bibinfo {author} {\bibfnamefont {P.}~\bibnamefont {Pani}},\ }\href@noop
  {} {\  (\bibinfo {year} {2020})},\ \Eprint {http://arxiv.org/abs/2004.00009}
  {arXiv:2004.00009 [gr-qc]} \BibitemShut {NoStop}%
\bibitem [{\citenamefont {Agathos}\ \emph {et~al.}(2014)\citenamefont
  {Agathos}, \citenamefont {Del~Pozzo}, \citenamefont {Li}, \citenamefont {Van
  Den~Broeck}, \citenamefont {Veitch},\ and\ \citenamefont
  {Vitale}}]{Agathos:2013upa}%
  \BibitemOpen
  \bibfield  {author} {\bibinfo {author} {\bibfnamefont {M.}~\bibnamefont
  {Agathos}}, \bibinfo {author} {\bibfnamefont {W.}~\bibnamefont {Del~Pozzo}},
  \bibinfo {author} {\bibfnamefont {T.~G.~F.}\ \bibnamefont {Li}}, \bibinfo
  {author} {\bibfnamefont {C.}~\bibnamefont {Van Den~Broeck}}, \bibinfo
  {author} {\bibfnamefont {J.}~\bibnamefont {Veitch}}, \ and\ \bibinfo {author}
  {\bibfnamefont {S.}~\bibnamefont {Vitale}},\ }\href {\doibase
  10.1103/PhysRevD.89.082001} {\bibfield  {journal} {\bibinfo  {journal}
  {Phys.\ Rev.\ D}\ }\textbf {\bibinfo {volume} {89}},\ \bibinfo {pages}
  {082001} (\bibinfo {year} {2014})},\ \Eprint {http://arxiv.org/abs/1311.0420}
  {arXiv:1311.0420 [gr-qc]} \BibitemShut {NoStop}%
\bibitem [{\citenamefont {Li}\ \emph {et~al.}(2012{\natexlab{a}})\citenamefont
  {Li}, \citenamefont {Del~Pozzo}, \citenamefont {Vitale}, \citenamefont {Van
  Den~Broeck}, \citenamefont {Agathos}, \citenamefont {Veitch}, \citenamefont
  {Grover}, \citenamefont {Sidery}, \citenamefont {Sturani},\ and\
  \citenamefont {Vecchio}}]{Li:2011cg}%
  \BibitemOpen
  \bibfield  {author} {\bibinfo {author} {\bibfnamefont {T.}~\bibnamefont
  {Li}}, \bibinfo {author} {\bibfnamefont {W.}~\bibnamefont {Del~Pozzo}},
  \bibinfo {author} {\bibfnamefont {S.}~\bibnamefont {Vitale}}, \bibinfo
  {author} {\bibfnamefont {C.}~\bibnamefont {Van Den~Broeck}}, \bibinfo
  {author} {\bibfnamefont {M.}~\bibnamefont {Agathos}}, \bibinfo {author}
  {\bibfnamefont {J.}~\bibnamefont {Veitch}}, \bibinfo {author} {\bibfnamefont
  {K.}~\bibnamefont {Grover}}, \bibinfo {author} {\bibfnamefont
  {T.}~\bibnamefont {Sidery}}, \bibinfo {author} {\bibfnamefont
  {R.}~\bibnamefont {Sturani}}, \ and\ \bibinfo {author} {\bibfnamefont
  {A.}~\bibnamefont {Vecchio}},\ }\href {\doibase 10.1103/PhysRevD.85.082003}
  {\bibfield  {journal} {\bibinfo  {journal} {Phys.\ Rev.\ D}\ }\textbf
  {\bibinfo {volume} {85}},\ \bibinfo {pages} {082003} (\bibinfo {year}
  {2012}{\natexlab{a}})},\ \Eprint {http://arxiv.org/abs/1110.0530}
  {arXiv:1110.0530 [gr-qc]} \BibitemShut {NoStop}%
\bibitem [{\citenamefont {Li}\ \emph {et~al.}(2012{\natexlab{b}})\citenamefont
  {Li}, \citenamefont {Del~Pozzo}, \citenamefont {Vitale}, \citenamefont {Van
  Den~Broeck}, \citenamefont {Agathos}, \citenamefont {Veitch}, \citenamefont
  {Grover}, \citenamefont {Sidery}, \citenamefont {Sturani},\ and\
  \citenamefont {Vecchio}}]{Li:2011vx}%
  \BibitemOpen
  \bibfield  {author} {\bibinfo {author} {\bibfnamefont {T.}~\bibnamefont
  {Li}}, \bibinfo {author} {\bibfnamefont {W.}~\bibnamefont {Del~Pozzo}},
  \bibinfo {author} {\bibfnamefont {S.}~\bibnamefont {Vitale}}, \bibinfo
  {author} {\bibfnamefont {C.}~\bibnamefont {Van Den~Broeck}}, \bibinfo
  {author} {\bibfnamefont {M.}~\bibnamefont {Agathos}}, \bibinfo {author}
  {\bibfnamefont {J.}~\bibnamefont {Veitch}}, \bibinfo {author} {\bibfnamefont
  {K.}~\bibnamefont {Grover}}, \bibinfo {author} {\bibfnamefont
  {T.}~\bibnamefont {Sidery}}, \bibinfo {author} {\bibfnamefont
  {R.}~\bibnamefont {Sturani}}, \ and\ \bibinfo {author} {\bibfnamefont
  {A.}~\bibnamefont {Vecchio}},\ }\href {\doibase
  10.1088/1742-6596/363/1/012028} {\bibfield  {journal} {\bibinfo  {journal}
  {J.\ Phys.\ Conf.\ Ser.}\ }\textbf {\bibinfo {volume} {363}},\ \bibinfo
  {pages} {012028} (\bibinfo {year} {2012}{\natexlab{b}})},\ \Eprint
  {http://arxiv.org/abs/1111.5274} {arXiv:1111.5274 [gr-qc]} \BibitemShut
  {NoStop}%
\bibitem [{\citenamefont {Cornish}\ \emph {et~al.}(2011)\citenamefont
  {Cornish}, \citenamefont {Sampson}, \citenamefont {Yunes},\ and\
  \citenamefont {Pretorius}}]{Cornish:2011ys}%
  \BibitemOpen
  \bibfield  {author} {\bibinfo {author} {\bibfnamefont {N.}~\bibnamefont
  {Cornish}}, \bibinfo {author} {\bibfnamefont {L.}~\bibnamefont {Sampson}},
  \bibinfo {author} {\bibfnamefont {N.}~\bibnamefont {Yunes}}, \ and\ \bibinfo
  {author} {\bibfnamefont {F.}~\bibnamefont {Pretorius}},\ }\href {\doibase
  10.1103/PhysRevD.84.062003} {\bibfield  {journal} {\bibinfo  {journal}
  {Phys.\ Rev.\ D}\ }\textbf {\bibinfo {volume} {84}},\ \bibinfo {pages}
  {062003} (\bibinfo {year} {2011})},\ \Eprint {http://arxiv.org/abs/1105.2088}
  {arXiv:1105.2088 [gr-qc]} \BibitemShut {NoStop}%
\bibitem [{\citenamefont {Meidam}\ \emph {et~al.}(2018)\citenamefont {Meidam}
  \emph {et~al.}}]{Meidam:2017dgf}%
  \BibitemOpen
  \bibfield  {author} {\bibinfo {author} {\bibfnamefont {J.}~\bibnamefont
  {Meidam}} \emph {et~al.},\ }\href {\doibase 10.1103/PhysRevD.97.044033}
  {\bibfield  {journal} {\bibinfo  {journal} {Phys.\ Rev.\ D}\ }\textbf
  {\bibinfo {volume} {97}},\ \bibinfo {pages} {044033} (\bibinfo {year}
  {2018})},\ \Eprint {http://arxiv.org/abs/1712.08772} {arXiv:1712.08772
  [gr-qc]} \BibitemShut {NoStop}%
\bibitem [{\citenamefont {Abbott}\ \emph
  {et~al.}(2016{\natexlab{b}})\citenamefont {Abbott} \emph
  {et~al.}}]{TheLIGOScientific:2016pea}%
  \BibitemOpen
  \bibfield  {author} {\bibinfo {author} {\bibfnamefont {B.}~\bibnamefont
  {Abbott}} \emph {et~al.} (\bibinfo {collaboration} {LIGO Scientific
  Collaboration, Virgo Collaboration}),\ }\href {\doibase
  10.1103/PhysRevX.6.041015} {\bibfield  {journal} {\bibinfo  {journal} {Phys.\
  Rev.\ X}\ }\textbf {\bibinfo {volume} {6}},\ \bibinfo {pages} {041015}
  (\bibinfo {year} {2016}{\natexlab{b}})},\ \bibinfo {note} {[Erratum:
  Phys.Rev.X 8, 039903 (2018)]},\ \Eprint {http://arxiv.org/abs/1606.04856}
  {arXiv:1606.04856 [gr-qc]} \BibitemShut {NoStop}%
\bibitem [{\citenamefont {Abbott}\ \emph {et~al.}(2017)\citenamefont {Abbott}
  \emph {et~al.}}]{Abbott:2017vtc}%
  \BibitemOpen
  \bibfield  {author} {\bibinfo {author} {\bibfnamefont {B.~P.}\ \bibnamefont
  {Abbott}} \emph {et~al.} (\bibinfo {collaboration} {LIGO Scientific
  Collaboration, Virgo Collaboration}),\ }\href {\doibase
  10.1103/PhysRevLett.118.221101} {\bibfield  {journal} {\bibinfo  {journal}
  {Phys.\ Rev.\ Lett.}\ }\textbf {\bibinfo {volume} {118}},\ \bibinfo {pages}
  {221101} (\bibinfo {year} {2017})},\ \bibinfo {note} {[Erratum:
  Phys.Rev.Lett. 121, 129901 (2018)]},\ \Eprint
  {http://arxiv.org/abs/1706.01812} {arXiv:1706.01812 [gr-qc]} \BibitemShut
  {NoStop}%
\bibitem [{\citenamefont {Chatziioannou}\ \emph {et~al.}(2012)\citenamefont
  {Chatziioannou}, \citenamefont {Yunes},\ and\ \citenamefont
  {Cornish}}]{Chatziioannou:2012rf}%
  \BibitemOpen
  \bibfield  {author} {\bibinfo {author} {\bibfnamefont {K.}~\bibnamefont
  {Chatziioannou}}, \bibinfo {author} {\bibfnamefont {N.}~\bibnamefont
  {Yunes}}, \ and\ \bibinfo {author} {\bibfnamefont {N.}~\bibnamefont
  {Cornish}},\ }\href {\doibase 10.1103/PhysRevD.86.022004} {\bibfield
  {journal} {\bibinfo  {journal} {Phys. Rev. D}\ }\textbf {\bibinfo {volume}
  {86}},\ \bibinfo {pages} {022004} (\bibinfo {year} {2012})},\ \bibinfo {note}
  {[Erratum: Phys.Rev.D 95, 129901 (2017)]},\ \Eprint
  {http://arxiv.org/abs/1204.2585} {arXiv:1204.2585 [gr-qc]} \BibitemShut
  {NoStop}%
\bibitem [{\citenamefont {Yunes}\ and\ \citenamefont
  {Siemens}(2013)}]{Yunes:2013dva}%
  \BibitemOpen
  \bibfield  {author} {\bibinfo {author} {\bibfnamefont {N.}~\bibnamefont
  {Yunes}}\ and\ \bibinfo {author} {\bibfnamefont {X.}~\bibnamefont
  {Siemens}},\ }\href {\doibase 10.12942/lrr-2013-9} {\bibfield  {journal}
  {\bibinfo  {journal} {Living Rev.\ Rel.}\ }\textbf {\bibinfo {volume} {16}},\
  \bibinfo {pages} {9} (\bibinfo {year} {2013})},\ \Eprint
  {http://arxiv.org/abs/1304.3473} {arXiv:1304.3473 [gr-qc]} \BibitemShut
  {NoStop}%
\bibitem [{\citenamefont {Yunes}\ \emph {et~al.}(2016)\citenamefont {Yunes},
  \citenamefont {Yagi},\ and\ \citenamefont {Pretorius}}]{Yunes:2016jcc}%
  \BibitemOpen
  \bibfield  {author} {\bibinfo {author} {\bibfnamefont {N.}~\bibnamefont
  {Yunes}}, \bibinfo {author} {\bibfnamefont {K.}~\bibnamefont {Yagi}}, \ and\
  \bibinfo {author} {\bibfnamefont {F.}~\bibnamefont {Pretorius}},\ }\href
  {\doibase 10.1103/PhysRevD.94.084002} {\bibfield  {journal} {\bibinfo
  {journal} {Phys.\ Rev.\ D}\ }\textbf {\bibinfo {volume} {94}},\ \bibinfo
  {pages} {084002} (\bibinfo {year} {2016})},\ \Eprint
  {http://arxiv.org/abs/1603.08955} {arXiv:1603.08955 [gr-qc]} \BibitemShut
  {NoStop}%
\bibitem [{\citenamefont {Nair}\ \emph {et~al.}(2019)\citenamefont {Nair},
  \citenamefont {Perkins}, \citenamefont {Silva},\ and\ \citenamefont
  {Yunes}}]{Nair:2019iur}%
  \BibitemOpen
  \bibfield  {author} {\bibinfo {author} {\bibfnamefont {R.}~\bibnamefont
  {Nair}}, \bibinfo {author} {\bibfnamefont {S.}~\bibnamefont {Perkins}},
  \bibinfo {author} {\bibfnamefont {H.~O.}\ \bibnamefont {Silva}}, \ and\
  \bibinfo {author} {\bibfnamefont {N.}~\bibnamefont {Yunes}},\ }\href
  {\doibase 10.1103/PhysRevLett.123.191101} {\bibfield  {journal} {\bibinfo
  {journal} {Phys. Rev. Lett.}\ }\textbf {\bibinfo {volume} {123}},\ \bibinfo
  {pages} {191101} (\bibinfo {year} {2019})},\ \Eprint
  {http://arxiv.org/abs/1905.00870} {arXiv:1905.00870 [gr-qc]} \BibitemShut
  {NoStop}%
\bibitem [{\citenamefont {Zimmerman}\ \emph {et~al.}(2019)\citenamefont
  {Zimmerman}, \citenamefont {Haster},\ and\ \citenamefont
  {Chatziioannou}}]{Zimmerman:2019wzo}%
  \BibitemOpen
  \bibfield  {author} {\bibinfo {author} {\bibfnamefont {A.}~\bibnamefont
  {Zimmerman}}, \bibinfo {author} {\bibfnamefont {C.-J.}\ \bibnamefont
  {Haster}}, \ and\ \bibinfo {author} {\bibfnamefont {K.}~\bibnamefont
  {Chatziioannou}},\ }\href {\doibase 10.1103/PhysRevD.99.124044} {\bibfield
  {journal} {\bibinfo  {journal} {Phys.\ Rev.\ D}\ }\textbf {\bibinfo {volume}
  {99}},\ \bibinfo {pages} {124044} (\bibinfo {year} {2019})},\ \Eprint
  {http://arxiv.org/abs/1903.11008} {arXiv:1903.11008 [astro-ph.IM]}
  \BibitemShut {NoStop}%
\bibitem [{\citenamefont {P{\"u}rrer}\ and\ \citenamefont
  {Haster}(2020)}]{Purrer:2019jcp}%
  \BibitemOpen
  \bibfield  {author} {\bibinfo {author} {\bibfnamefont {M.}~\bibnamefont
  {P{\"u}rrer}}\ and\ \bibinfo {author} {\bibfnamefont {C.-J.}\ \bibnamefont
  {Haster}},\ }\href {\doibase 10.1103/PhysRevResearch.2.023151} {\bibfield
  {journal} {\bibinfo  {journal} {Phys.\ Rev.\ Research}\ }\textbf {\bibinfo
  {volume} {2}},\ \bibinfo {pages} {023151} (\bibinfo {year} {2020})},\ \Eprint
  {http://arxiv.org/abs/1912.10055} {arXiv:1912.10055 [gr-qc]} \BibitemShut
  {NoStop}%
\bibitem [{\citenamefont {Thorne}(1980)}]{Thorne:1980ru}%
  \BibitemOpen
  \bibfield  {author} {\bibinfo {author} {\bibfnamefont {K.}~\bibnamefont
  {Thorne}},\ }\href {\doibase 10.1103/RevModPhys.52.299} {\bibfield  {journal}
  {\bibinfo  {journal} {Rev. Mod. Phys.}\ }\textbf {\bibinfo {volume} {52}},\
  \bibinfo {pages} {299} (\bibinfo {year} {1980})}\BibitemShut {NoStop}%
\bibitem [{\citenamefont {Blanchet}\ and\ \citenamefont
  {Damour}(1988)}]{Blanchet:1987wq}%
  \BibitemOpen
  \bibfield  {author} {\bibinfo {author} {\bibfnamefont {L.}~\bibnamefont
  {Blanchet}}\ and\ \bibinfo {author} {\bibfnamefont {T.}~\bibnamefont
  {Damour}},\ }\href {\doibase 10.1103/PhysRevD.37.1410} {\bibfield  {journal}
  {\bibinfo  {journal} {Phys. Rev. D}\ }\textbf {\bibinfo {volume} {37}},\
  \bibinfo {pages} {1410} (\bibinfo {year} {1988})}\BibitemShut {NoStop}%
\bibitem [{\citenamefont {Blanchet}\ and\ \citenamefont
  {Damour}(1992)}]{Blanchet:1992br}%
  \BibitemOpen
  \bibfield  {author} {\bibinfo {author} {\bibfnamefont {L.}~\bibnamefont
  {Blanchet}}\ and\ \bibinfo {author} {\bibfnamefont {T.}~\bibnamefont
  {Damour}},\ }\href {\doibase 10.1103/PhysRevD.46.4304} {\bibfield  {journal}
  {\bibinfo  {journal} {Phys. Rev. D}\ }\textbf {\bibinfo {volume} {46}},\
  \bibinfo {pages} {4304} (\bibinfo {year} {1992})}\BibitemShut {NoStop}%
\bibitem [{\citenamefont {Blanchet}\ and\ \citenamefont
  {Sch{\"a}fer}(1993)}]{Blanchet:1993ec}%
  \BibitemOpen
  \bibfield  {author} {\bibinfo {author} {\bibfnamefont {L.}~\bibnamefont
  {Blanchet}}\ and\ \bibinfo {author} {\bibfnamefont {G.}~\bibnamefont
  {Sch{\"a}fer}},\ }\href {\doibase 10.1088/0264-9381/10/12/026} {\bibfield
  {journal} {\bibinfo  {journal} {Class.\ Quant.\ Grav.}\ }\textbf {\bibinfo
  {volume} {10}},\ \bibinfo {pages} {2699} (\bibinfo {year}
  {1993})}\BibitemShut {NoStop}%
\bibitem [{\citenamefont {Tanaka}\ \emph {et~al.}(1993)\citenamefont {Tanaka},
  \citenamefont {Shibata}, \citenamefont {Sasaki}, \citenamefont {Tagoshi},\
  and\ \citenamefont {Nakamura}}]{Tanaka:1993pu}%
  \BibitemOpen
  \bibfield  {author} {\bibinfo {author} {\bibfnamefont {T.}~\bibnamefont
  {Tanaka}}, \bibinfo {author} {\bibfnamefont {M.}~\bibnamefont {Shibata}},
  \bibinfo {author} {\bibfnamefont {M.}~\bibnamefont {Sasaki}}, \bibinfo
  {author} {\bibfnamefont {H.}~\bibnamefont {Tagoshi}}, \ and\ \bibinfo
  {author} {\bibfnamefont {T.}~\bibnamefont {Nakamura}},\ }\href {\doibase
  10.1143/PTP.90.65} {\bibfield  {journal} {\bibinfo  {journal} {Prog.\ Theor.\
  Phys.}\ }\textbf {\bibinfo {volume} {90}},\ \bibinfo {pages} {65} (\bibinfo
  {year} {1993})}\BibitemShut {NoStop}%
\bibitem [{\citenamefont {Blanchet}\ and\ \citenamefont
  {Sathyaprakash}(1995)}]{Blanchet:1994ez}%
  \BibitemOpen
  \bibfield  {author} {\bibinfo {author} {\bibfnamefont {L.}~\bibnamefont
  {Blanchet}}\ and\ \bibinfo {author} {\bibfnamefont {B.}~\bibnamefont
  {Sathyaprakash}},\ }\href {\doibase 10.1103/PhysRevLett.74.1067} {\bibfield
  {journal} {\bibinfo  {journal} {Phys.\ Rev.\ Lett.}\ }\textbf {\bibinfo
  {volume} {74}},\ \bibinfo {pages} {1067} (\bibinfo {year}
  {1995})}\BibitemShut {NoStop}%
\bibitem [{\citenamefont {Blanchet}(1998)}]{Blanchet:1997jj}%
  \BibitemOpen
  \bibfield  {author} {\bibinfo {author} {\bibfnamefont {L.}~\bibnamefont
  {Blanchet}},\ }\href {\doibase 10.1088/0264-9381/15/1/009} {\bibfield
  {journal} {\bibinfo  {journal} {Class.\ Quant.\ Grav.}\ }\textbf {\bibinfo
  {volume} {15}},\ \bibinfo {pages} {113} (\bibinfo {year} {1998})},\ \bibinfo
  {note} {[Erratum: Class.Quant.Grav. 22, 3381 (2005)]},\ \Eprint
  {http://arxiv.org/abs/gr-qc/9710038} {arXiv:gr-qc/9710038} \BibitemShut
  {NoStop}%
\bibitem [{\citenamefont {\mbox{The On-Line Encyclopedia of Integer
  Sequences}}(2020)}]{OEISpi}%
  \BibitemOpen
  \bibfield  {author} {\bibinfo {author} {\bibnamefont {\mbox{The On-Line
  Encyclopedia of Integer Sequences}}},\ }\href {https://oeis.org/A000796}
  {\enquote {\bibinfo {title} {{Sequence A000108 -- Decimal expansion of
  Pi}},}\ } (\bibinfo {year} {2020})\BibitemShut {NoStop}%
\bibitem [{\citenamefont {Ramaley}(1969)}]{Ramaley1969}%
  \BibitemOpen
  \bibfield  {author} {\bibinfo {author} {\bibfnamefont {J.~F.}\ \bibnamefont
  {Ramaley}},\ }\href {\doibase 10.2307/2317945} {\bibfield  {journal}
  {\bibinfo  {journal} {The American Mathematical Monthly}\ }\textbf {\bibinfo
  {volume} {76}},\ \bibinfo {pages} {916} (\bibinfo {year} {1969})}\BibitemShut
  {NoStop}%
\bibitem [{\citenamefont {Bailey}\ \emph {et~al.}(1997)\citenamefont {Bailey},
  \citenamefont {Borwein},\ and\ \citenamefont {Plouffe}}]{Bailey1997}%
  \BibitemOpen
  \bibfield  {author} {\bibinfo {author} {\bibfnamefont {D.}~\bibnamefont
  {Bailey}}, \bibinfo {author} {\bibfnamefont {P.}~\bibnamefont {Borwein}}, \
  and\ \bibinfo {author} {\bibfnamefont {S.}~\bibnamefont {Plouffe}},\ }\href
  {\doibase 10.1090/s0025-5718-97-00856-9} {\bibfield  {journal} {\bibinfo
  {journal} {Mathematics of Computation}\ }\textbf {\bibinfo {volume} {66}},\
  \bibinfo {pages} {903} (\bibinfo {year} {1997})}\BibitemShut {NoStop}%
\bibitem [{\citenamefont {Galperin}(2003)}]{Galperin2003}%
  \BibitemOpen
  \bibfield  {author} {\bibinfo {author} {\bibfnamefont {G.}~\bibnamefont
  {Galperin}},\ }\href {\doibase 10.1070/rd2003v008n04abeh000252} {\bibfield
  {journal} {\bibinfo  {journal} {Regular and Chaotic Dynamics}\ }\textbf
  {\bibinfo {volume} {8}},\ \bibinfo {pages} {375} (\bibinfo {year}
  {2003})}\BibitemShut {NoStop}%
\bibitem [{\citenamefont {{Dumoulin}}\ and\ \citenamefont
  {{Thouin}}(2014)}]{Dumoulin:2014}%
  \BibitemOpen
  \bibfield  {author} {\bibinfo {author} {\bibfnamefont {V.}~\bibnamefont
  {{Dumoulin}}}\ and\ \bibinfo {author} {\bibfnamefont {F.}~\bibnamefont
  {{Thouin}}},\ }\href@noop {} {\bibfield  {journal} {\bibinfo  {journal}
  {ArXiv e-prints}\ } (\bibinfo {year} {2014})},\ \Eprint
  {http://arxiv.org/abs/1404.1499} {arXiv:1404.1499 [physics.pop-ph]}
  \BibitemShut {NoStop}%
\bibitem [{\citenamefont {Arndt}\ and\ \citenamefont
  {Haenel}(2001)}]{Arndt2001}%
  \BibitemOpen
  \bibfield  {author} {\bibinfo {author} {\bibfnamefont {J.}~\bibnamefont
  {Arndt}}\ and\ \bibinfo {author} {\bibfnamefont {C.}~\bibnamefont {Haenel}},\
  }\href {\doibase 10.1007/978-3-642-56735-3} {\emph {\bibinfo {title} {Pi
  {\textemdash} Unleashed}}}\ (\bibinfo  {publisher} {Springer Berlin
  Heidelberg},\ \bibinfo {year} {2001})\BibitemShut {NoStop}%
\bibitem [{\citenamefont {Yee}(2020)}]{yCruncher}%
  \BibitemOpen
  \bibfield  {author} {\bibinfo {author} {\bibfnamefont {A.~E.}\ \bibnamefont
  {Yee}},\ }\href {http://www.numberworld.org/y-cruncher/} {\enquote {\bibinfo
  {title} {{y-cruncher - A Multi-Threaded Pi-Program}},}\ } (\bibinfo {year}
  {2020})\BibitemShut {NoStop}%
\bibitem [{\citenamefont {Sathyaprakash}\ and\ \citenamefont
  {Dhurandhar}(1991)}]{Sathyaprakash:1991mt}%
  \BibitemOpen
  \bibfield  {author} {\bibinfo {author} {\bibfnamefont {B.}~\bibnamefont
  {Sathyaprakash}}\ and\ \bibinfo {author} {\bibfnamefont {S.}~\bibnamefont
  {Dhurandhar}},\ }\href {\doibase 10.1103/PhysRevD.44.3819} {\bibfield
  {journal} {\bibinfo  {journal} {Phys. Rev. D}\ }\textbf {\bibinfo {volume}
  {44}},\ \bibinfo {pages} {3819} (\bibinfo {year} {1991})}\BibitemShut
  {NoStop}%
\bibitem [{\citenamefont {Cutler}\ and\ \citenamefont
  {Flanagan}(1994)}]{Cutler:1994ys}%
  \BibitemOpen
  \bibfield  {author} {\bibinfo {author} {\bibfnamefont {C.}~\bibnamefont
  {Cutler}}\ and\ \bibinfo {author} {\bibfnamefont {{\'E}.~{\'E}.}\
  \bibnamefont {Flanagan}},\ }\href {\doibase 10.1103/PhysRevD.49.2658}
  {\bibfield  {journal} {\bibinfo  {journal} {Phys.\ Rev.\ D}\ }\textbf
  {\bibinfo {volume} {49}},\ \bibinfo {pages} {2658} (\bibinfo {year}
  {1994})},\ \Eprint {http://arxiv.org/abs/gr-qc/9402014} {arXiv:gr-qc/9402014}
  \BibitemShut {NoStop}%
\bibitem [{\citenamefont {Apostolatos}\ \emph {et~al.}(1994)\citenamefont
  {Apostolatos}, \citenamefont {Cutler}, \citenamefont {Sussman},\ and\
  \citenamefont {Thorne}}]{Apostolatos:1994mx}%
  \BibitemOpen
  \bibfield  {author} {\bibinfo {author} {\bibfnamefont {T.~A.}\ \bibnamefont
  {Apostolatos}}, \bibinfo {author} {\bibfnamefont {C.}~\bibnamefont {Cutler}},
  \bibinfo {author} {\bibfnamefont {G.~J.}\ \bibnamefont {Sussman}}, \ and\
  \bibinfo {author} {\bibfnamefont {K.~S.}\ \bibnamefont {Thorne}},\ }\href
  {\doibase 10.1103/PhysRevD.49.6274} {\bibfield  {journal} {\bibinfo
  {journal} {Phys.\ Rev.\ D}\ }\textbf {\bibinfo {volume} {49}},\ \bibinfo
  {pages} {6274} (\bibinfo {year} {1994})}\BibitemShut {NoStop}%
\bibitem [{\citenamefont {Poisson}\ and\ \citenamefont
  {Will}(1995)}]{Poisson:1995ef}%
  \BibitemOpen
  \bibfield  {author} {\bibinfo {author} {\bibfnamefont {E.}~\bibnamefont
  {Poisson}}\ and\ \bibinfo {author} {\bibfnamefont {C.~M.}\ \bibnamefont
  {Will}},\ }\href {\doibase 10.1103/PhysRevD.52.848} {\bibfield  {journal}
  {\bibinfo  {journal} {Phys.\ Rev.\ D}\ }\textbf {\bibinfo {volume} {52}},\
  \bibinfo {pages} {848} (\bibinfo {year} {1995})},\ \Eprint
  {http://arxiv.org/abs/gr-qc/9502040} {arXiv:gr-qc/9502040} \BibitemShut
  {NoStop}%
\bibitem [{\citenamefont {Droz}\ \emph {et~al.}(1999)\citenamefont {Droz},
  \citenamefont {Knapp}, \citenamefont {Poisson},\ and\ \citenamefont
  {Owen}}]{Droz:1999qx}%
  \BibitemOpen
  \bibfield  {author} {\bibinfo {author} {\bibfnamefont {S.}~\bibnamefont
  {Droz}}, \bibinfo {author} {\bibfnamefont {D.~J.}\ \bibnamefont {Knapp}},
  \bibinfo {author} {\bibfnamefont {E.}~\bibnamefont {Poisson}}, \ and\
  \bibinfo {author} {\bibfnamefont {B.~J.}\ \bibnamefont {Owen}},\ }\href
  {\doibase 10.1103/PhysRevD.59.124016} {\bibfield  {journal} {\bibinfo
  {journal} {Phys.\ Rev.\ D}\ }\textbf {\bibinfo {volume} {59}},\ \bibinfo
  {pages} {124016} (\bibinfo {year} {1999})},\ \Eprint
  {http://arxiv.org/abs/gr-qc/9901076} {arXiv:gr-qc/9901076} \BibitemShut
  {NoStop}%
\bibitem [{\citenamefont {Buonanno}\ \emph {et~al.}(2009)\citenamefont
  {Buonanno}, \citenamefont {Iyer}, \citenamefont {Ochsner}, \citenamefont
  {Pan},\ and\ \citenamefont {Sathyaprakash}}]{Buonanno:2009zt}%
  \BibitemOpen
  \bibfield  {author} {\bibinfo {author} {\bibfnamefont {A.}~\bibnamefont
  {Buonanno}}, \bibinfo {author} {\bibfnamefont {B.}~\bibnamefont {Iyer}},
  \bibinfo {author} {\bibfnamefont {E.}~\bibnamefont {Ochsner}}, \bibinfo
  {author} {\bibfnamefont {Y.}~\bibnamefont {Pan}}, \ and\ \bibinfo {author}
  {\bibfnamefont {B.}~\bibnamefont {Sathyaprakash}},\ }\href {\doibase
  10.1103/PhysRevD.80.084043} {\bibfield  {journal} {\bibinfo  {journal}
  {Phys.\ Rev.\ D}\ }\textbf {\bibinfo {volume} {80}},\ \bibinfo {pages}
  {084043} (\bibinfo {year} {2009})},\ \Eprint {http://arxiv.org/abs/0907.0700}
  {arXiv:0907.0700 [gr-qc]} \BibitemShut {NoStop}%
\bibitem [{\citenamefont {Damour}\ \emph {et~al.}(2001)\citenamefont {Damour},
  \citenamefont {Iyer},\ and\ \citenamefont {Sathyaprakash}}]{Damour:2000zb}%
  \BibitemOpen
  \bibfield  {author} {\bibinfo {author} {\bibfnamefont {T.}~\bibnamefont
  {Damour}}, \bibinfo {author} {\bibfnamefont {B.~R.}\ \bibnamefont {Iyer}}, \
  and\ \bibinfo {author} {\bibfnamefont {B.}~\bibnamefont {Sathyaprakash}},\
  }\href {\doibase 10.1103/PhysRevD.63.044023} {\bibfield  {journal} {\bibinfo
  {journal} {Phys.\ Rev.\ D}\ }\textbf {\bibinfo {volume} {63}},\ \bibinfo
  {pages} {044023} (\bibinfo {year} {2001})},\ \bibinfo {note} {[Erratum:
  Phys.Rev.D 72, 029902 (2005)]},\ \Eprint {http://arxiv.org/abs/gr-qc/0010009}
  {arXiv:gr-qc/0010009} \BibitemShut {NoStop}%
\bibitem [{\citenamefont {Damour}\ \emph {et~al.}(2002)\citenamefont {Damour},
  \citenamefont {Iyer},\ and\ \citenamefont {Sathyaprakash}}]{Damour:2002kr}%
  \BibitemOpen
  \bibfield  {author} {\bibinfo {author} {\bibfnamefont {T.}~\bibnamefont
  {Damour}}, \bibinfo {author} {\bibfnamefont {B.~R.}\ \bibnamefont {Iyer}}, \
  and\ \bibinfo {author} {\bibfnamefont {B.}~\bibnamefont {Sathyaprakash}},\
  }\href {\doibase 10.1103/PhysRevD.66.027502} {\bibfield  {journal} {\bibinfo
  {journal} {Phys.\ Rev.\ D}\ }\textbf {\bibinfo {volume} {66}},\ \bibinfo
  {pages} {027502} (\bibinfo {year} {2002})},\ \Eprint
  {http://arxiv.org/abs/gr-qc/0207021} {arXiv:gr-qc/0207021} \BibitemShut
  {NoStop}%
\bibitem [{\citenamefont {Arun}\ \emph {et~al.}(2005)\citenamefont {Arun},
  \citenamefont {Iyer}, \citenamefont {Sathyaprakash},\ and\ \citenamefont
  {Sundararajan}}]{Arun:2004hn}%
  \BibitemOpen
  \bibfield  {author} {\bibinfo {author} {\bibfnamefont {K.}~\bibnamefont
  {Arun}}, \bibinfo {author} {\bibfnamefont {B.~R.}\ \bibnamefont {Iyer}},
  \bibinfo {author} {\bibfnamefont {B.}~\bibnamefont {Sathyaprakash}}, \ and\
  \bibinfo {author} {\bibfnamefont {P.~A.}\ \bibnamefont {Sundararajan}},\
  }\href {\doibase 10.1103/PhysRevD.71.084008} {\bibfield  {journal} {\bibinfo
  {journal} {Phys.\ Rev.\ D}\ }\textbf {\bibinfo {volume} {71}},\ \bibinfo
  {pages} {084008} (\bibinfo {year} {2005})},\ \bibinfo {note} {[Erratum:
  Phys.Rev.D 72, 069903 (2005)]},\ \Eprint {http://arxiv.org/abs/gr-qc/0411146}
  {arXiv:gr-qc/0411146} \BibitemShut {NoStop}%
\bibitem [{\citenamefont {Boh{\'e}}\ \emph {et~al.}(2013)\citenamefont
  {Boh{\'e}}, \citenamefont {Marsat},\ and\ \citenamefont
  {Blanchet}}]{Bohe:2013cla}%
  \BibitemOpen
  \bibfield  {author} {\bibinfo {author} {\bibfnamefont {A.}~\bibnamefont
  {Boh{\'e}}}, \bibinfo {author} {\bibfnamefont {S.}~\bibnamefont {Marsat}}, \
  and\ \bibinfo {author} {\bibfnamefont {L.}~\bibnamefont {Blanchet}},\ }\href
  {\doibase 10.1088/0264-9381/30/13/135009} {\bibfield  {journal} {\bibinfo
  {journal} {Class.\ Quant.\ Grav.}\ }\textbf {\bibinfo {volume} {30}},\
  \bibinfo {pages} {135009} (\bibinfo {year} {2013})},\ \Eprint
  {http://arxiv.org/abs/1303.7412} {arXiv:1303.7412 [gr-qc]} \BibitemShut
  {NoStop}%
\bibitem [{\citenamefont {Poisson}(1998)}]{Poisson:1997ha}%
  \BibitemOpen
  \bibfield  {author} {\bibinfo {author} {\bibfnamefont {E.}~\bibnamefont
  {Poisson}},\ }\href {\doibase 10.1103/PhysRevD.57.5287} {\bibfield  {journal}
  {\bibinfo  {journal} {Phys.\ Rev.\ D}\ }\textbf {\bibinfo {volume} {57}},\
  \bibinfo {pages} {5287} (\bibinfo {year} {1998})},\ \Eprint
  {http://arxiv.org/abs/gr-qc/9709032} {arXiv:gr-qc/9709032} \BibitemShut
  {NoStop}%
\bibitem [{\citenamefont {Arun}\ \emph {et~al.}(2009)\citenamefont {Arun},
  \citenamefont {Buonanno}, \citenamefont {Faye},\ and\ \citenamefont
  {Ochsner}}]{Arun:2008kb}%
  \BibitemOpen
  \bibfield  {author} {\bibinfo {author} {\bibfnamefont {K.}~\bibnamefont
  {Arun}}, \bibinfo {author} {\bibfnamefont {A.}~\bibnamefont {Buonanno}},
  \bibinfo {author} {\bibfnamefont {G.}~\bibnamefont {Faye}}, \ and\ \bibinfo
  {author} {\bibfnamefont {E.}~\bibnamefont {Ochsner}},\ }\href {\doibase
  10.1103/PhysRevD.79.104023} {\bibfield  {journal} {\bibinfo  {journal}
  {Phys.\ Rev.\ D}\ }\textbf {\bibinfo {volume} {79}},\ \bibinfo {pages}
  {104023} (\bibinfo {year} {2009})},\ \bibinfo {note} {[Erratum: Phys.Rev.D
  84, 049901 (2011)]},\ \Eprint {http://arxiv.org/abs/0810.5336}
  {arXiv:0810.5336 [gr-qc]} \BibitemShut {NoStop}%
\bibitem [{\citenamefont {Mikoczi}\ \emph {et~al.}(2005)\citenamefont
  {Mikoczi}, \citenamefont {Vasuth},\ and\ \citenamefont
  {Gergely}}]{Mikoczi:2005dn}%
  \BibitemOpen
  \bibfield  {author} {\bibinfo {author} {\bibfnamefont {B.}~\bibnamefont
  {Mikoczi}}, \bibinfo {author} {\bibfnamefont {M.}~\bibnamefont {Vasuth}}, \
  and\ \bibinfo {author} {\bibfnamefont {L.~A.}\ \bibnamefont {Gergely}},\
  }\href {\doibase 10.1103/PhysRevD.71.124043} {\bibfield  {journal} {\bibinfo
  {journal} {Phys.\ Rev.\ D}\ }\textbf {\bibinfo {volume} {71}},\ \bibinfo
  {pages} {124043} (\bibinfo {year} {2005})},\ \Eprint
  {http://arxiv.org/abs/astro-ph/0504538} {arXiv:astro-ph/0504538} \BibitemShut
  {NoStop}%
\bibitem [{\citenamefont {Khan}\ \emph {et~al.}(2016)\citenamefont {Khan},
  \citenamefont {Husa}, \citenamefont {Hannam}, \citenamefont {Ohme},
  \citenamefont {P{\"u}rrer}, \citenamefont {Jim{\'e}nez~Forteza},\ and\
  \citenamefont {Boh{\'e}}}]{Khan:2015jqa}%
  \BibitemOpen
  \bibfield  {author} {\bibinfo {author} {\bibfnamefont {S.}~\bibnamefont
  {Khan}}, \bibinfo {author} {\bibfnamefont {S.}~\bibnamefont {Husa}}, \bibinfo
  {author} {\bibfnamefont {M.}~\bibnamefont {Hannam}}, \bibinfo {author}
  {\bibfnamefont {F.}~\bibnamefont {Ohme}}, \bibinfo {author} {\bibfnamefont
  {M.}~\bibnamefont {P{\"u}rrer}}, \bibinfo {author} {\bibfnamefont
  {X.}~\bibnamefont {Jim{\'e}nez~Forteza}}, \ and\ \bibinfo {author}
  {\bibfnamefont {A.}~\bibnamefont {Boh{\'e}}},\ }\href {\doibase
  10.1103/PhysRevD.93.044007} {\bibfield  {journal} {\bibinfo  {journal}
  {Phys.\ Rev.\ D}\ }\textbf {\bibinfo {volume} {93}},\ \bibinfo {pages}
  {044007} (\bibinfo {year} {2016})},\ \Eprint
  {http://arxiv.org/abs/1508.07253} {arXiv:1508.07253 [gr-qc]} \BibitemShut
  {NoStop}%
\bibitem [{\citenamefont {Husa}\ \emph {et~al.}(2016)\citenamefont {Husa},
  \citenamefont {Khan}, \citenamefont {Hannam}, \citenamefont {P{\"u}rrer},
  \citenamefont {Ohme}, \citenamefont {Jim{\'e}nez~Forteza},\ and\
  \citenamefont {Boh{\'e}}}]{Husa:2015iqa}%
  \BibitemOpen
  \bibfield  {author} {\bibinfo {author} {\bibfnamefont {S.}~\bibnamefont
  {Husa}}, \bibinfo {author} {\bibfnamefont {S.}~\bibnamefont {Khan}}, \bibinfo
  {author} {\bibfnamefont {M.}~\bibnamefont {Hannam}}, \bibinfo {author}
  {\bibfnamefont {M.}~\bibnamefont {P{\"u}rrer}}, \bibinfo {author}
  {\bibfnamefont {F.}~\bibnamefont {Ohme}}, \bibinfo {author} {\bibfnamefont
  {X.}~\bibnamefont {Jim{\'e}nez~Forteza}}, \ and\ \bibinfo {author}
  {\bibfnamefont {A.}~\bibnamefont {Boh{\'e}}},\ }\href {\doibase
  10.1103/PhysRevD.93.044006} {\bibfield  {journal} {\bibinfo  {journal}
  {Phys.\ Rev.\ D}\ }\textbf {\bibinfo {volume} {93}},\ \bibinfo {pages}
  {044006} (\bibinfo {year} {2016})},\ \Eprint
  {http://arxiv.org/abs/1508.07250} {arXiv:1508.07250 [gr-qc]} \BibitemShut
  {NoStop}%
\bibitem [{\citenamefont {Hannam}\ \emph {et~al.}(2014)\citenamefont {Hannam},
  \citenamefont {Schmidt}, \citenamefont {Boh{\'e}}, \citenamefont {Haegel},
  \citenamefont {Husa}, \citenamefont {Ohme}, \citenamefont {Pratten},\ and\
  \citenamefont {P{\"u}rrer}}]{Hannam:2013oca}%
  \BibitemOpen
  \bibfield  {author} {\bibinfo {author} {\bibfnamefont {M.}~\bibnamefont
  {Hannam}}, \bibinfo {author} {\bibfnamefont {P.}~\bibnamefont {Schmidt}},
  \bibinfo {author} {\bibfnamefont {A.}~\bibnamefont {Boh{\'e}}}, \bibinfo
  {author} {\bibfnamefont {L.}~\bibnamefont {Haegel}}, \bibinfo {author}
  {\bibfnamefont {S.}~\bibnamefont {Husa}}, \bibinfo {author} {\bibfnamefont
  {F.}~\bibnamefont {Ohme}}, \bibinfo {author} {\bibfnamefont {G.}~\bibnamefont
  {Pratten}}, \ and\ \bibinfo {author} {\bibfnamefont {M.}~\bibnamefont
  {P{\"u}rrer}},\ }\href {\doibase 10.1103/PhysRevLett.113.151101} {\bibfield
  {journal} {\bibinfo  {journal} {Phys.\ Rev.\ Lett.}\ }\textbf {\bibinfo
  {volume} {113}},\ \bibinfo {pages} {151101} (\bibinfo {year} {2014})},\
  \Eprint {http://arxiv.org/abs/1308.3271} {arXiv:1308.3271 [gr-qc]}
  \BibitemShut {NoStop}%
\bibitem [{\citenamefont {Dietrich}\ \emph {et~al.}(2019)\citenamefont
  {Dietrich} \emph {et~al.}}]{Dietrich:2018uni}%
  \BibitemOpen
  \bibfield  {author} {\bibinfo {author} {\bibfnamefont {T.}~\bibnamefont
  {Dietrich}} \emph {et~al.},\ }\href {\doibase 10.1103/PhysRevD.99.024029}
  {\bibfield  {journal} {\bibinfo  {journal} {Phys.\ Rev.\ D}\ }\textbf
  {\bibinfo {volume} {99}},\ \bibinfo {pages} {024029} (\bibinfo {year}
  {2019})},\ \Eprint {http://arxiv.org/abs/1804.02235} {arXiv:1804.02235
  [gr-qc]} \BibitemShut {NoStop}%
\bibitem [{\citenamefont {Dietrich}\ \emph {et~al.}(2017)\citenamefont
  {Dietrich}, \citenamefont {Bernuzzi},\ and\ \citenamefont
  {Tichy}}]{Dietrich:2017aum}%
  \BibitemOpen
  \bibfield  {author} {\bibinfo {author} {\bibfnamefont {T.}~\bibnamefont
  {Dietrich}}, \bibinfo {author} {\bibfnamefont {S.}~\bibnamefont {Bernuzzi}},
  \ and\ \bibinfo {author} {\bibfnamefont {W.}~\bibnamefont {Tichy}},\ }\href
  {\doibase 10.1103/PhysRevD.96.121501} {\bibfield  {journal} {\bibinfo
  {journal} {Phys.\ Rev.\ D}\ }\textbf {\bibinfo {volume} {96}},\ \bibinfo
  {pages} {121501} (\bibinfo {year} {2017})},\ \Eprint
  {http://arxiv.org/abs/1706.02969} {arXiv:1706.02969 [gr-qc]} \BibitemShut
  {NoStop}%
\bibitem [{\citenamefont {Foffa}\ and\ \citenamefont
  {Sturani}(2011)}]{Foffa:2011ub}%
  \BibitemOpen
  \bibfield  {author} {\bibinfo {author} {\bibfnamefont {S.}~\bibnamefont
  {Foffa}}\ and\ \bibinfo {author} {\bibfnamefont {R.}~\bibnamefont
  {Sturani}},\ }\href {\doibase 10.1103/PhysRevD.84.044031} {\bibfield
  {journal} {\bibinfo  {journal} {Phys.\ Rev.\ D}\ }\textbf {\bibinfo {volume}
  {84}},\ \bibinfo {pages} {044031} (\bibinfo {year} {2011})},\ \Eprint
  {http://arxiv.org/abs/1104.1122} {arXiv:1104.1122 [gr-qc]} \BibitemShut
  {NoStop}%
\bibitem [{\citenamefont {Veitch}\ \emph {et~al.}(2015)\citenamefont {Veitch}
  \emph {et~al.}}]{Veitch:2014wba}%
  \BibitemOpen
  \bibfield  {author} {\bibinfo {author} {\bibfnamefont {J.}~\bibnamefont
  {Veitch}} \emph {et~al.},\ }\href {\doibase 10.1103/PhysRevD.91.042003}
  {\bibfield  {journal} {\bibinfo  {journal} {Phys.\ Rev.\ D}\ }\textbf
  {\bibinfo {volume} {91}},\ \bibinfo {pages} {042003} (\bibinfo {year}
  {2015})},\ \Eprint {http://arxiv.org/abs/1409.7215} {arXiv:1409.7215 [gr-qc]}
  \BibitemShut {NoStop}%
\bibitem [{LAL(2019)}]{LALInference-code}%
  \BibitemOpen
  \href@noop {} {\enquote {\bibinfo {title} {\textsc{LALInference}},}\
  }\bibinfo {howpublished}
  {\url{https://git.ligo.org/lscsoft/lalsuite/tree/master/lalinference}}
  (\bibinfo {year} {2019})\BibitemShut {NoStop}%
\bibitem [{\citenamefont {{LIGO Scientific Collaboration}}(2019)}]{lalsuite}%
  \BibitemOpen
  \bibfield  {author} {\bibinfo {author} {\bibnamefont {{LIGO Scientific
  Collaboration}}},\ }\href {\doibase 10.7935/GT1W-FZ16} {\enquote {\bibinfo
  {title} {{LIGO} {A}lgorithm {L}ibrary - {LALS}uite},}\ }\bibinfo
  {howpublished} {free software (GPL)} (\bibinfo {year} {2019})\BibitemShut
  {NoStop}%
\bibitem [{\citenamefont {Abbott}\ \emph
  {et~al.}(2019{\natexlab{c}})\citenamefont {Abbott} \emph
  {et~al.}}]{LIGOScientific:2018mvr}%
  \BibitemOpen
  \bibfield  {author} {\bibinfo {author} {\bibfnamefont {B.}~\bibnamefont
  {Abbott}} \emph {et~al.} (\bibinfo {collaboration} {LIGO Scientific
  Collaboration, Virgo Collaboration}),\ }\href {\doibase
  10.1103/PhysRevX.9.031040} {\bibfield  {journal} {\bibinfo  {journal} {Phys.\
  Rev.\ X}\ }\textbf {\bibinfo {volume} {9}},\ \bibinfo {pages} {031040}
  (\bibinfo {year} {2019}{\natexlab{c}})},\ \Eprint
  {http://arxiv.org/abs/1811.12907} {arXiv:1811.12907 [astro-ph.HE]}
  \BibitemShut {NoStop}%
\bibitem [{\citenamefont {Littenberg}\ and\ \citenamefont
  {Cornish}(2015)}]{Littenberg:2014oda}%
  \BibitemOpen
  \bibfield  {author} {\bibinfo {author} {\bibfnamefont {T.~B.}\ \bibnamefont
  {Littenberg}}\ and\ \bibinfo {author} {\bibfnamefont {N.~J.}\ \bibnamefont
  {Cornish}},\ }\href {\doibase 10.1103/PhysRevD.91.084034} {\bibfield
  {journal} {\bibinfo  {journal} {Phys.\ Rev.\ D}\ }\textbf {\bibinfo {volume}
  {91}},\ \bibinfo {pages} {084034} (\bibinfo {year} {2015})},\ \Eprint
  {http://arxiv.org/abs/1410.3852} {arXiv:1410.3852 [gr-qc]} \BibitemShut
  {NoStop}%
\bibitem [{\citenamefont {Cornish}\ and\ \citenamefont
  {Littenberg}(2015)}]{Cornish:2014kda}%
  \BibitemOpen
  \bibfield  {author} {\bibinfo {author} {\bibfnamefont {N.~J.}\ \bibnamefont
  {Cornish}}\ and\ \bibinfo {author} {\bibfnamefont {T.~B.}\ \bibnamefont
  {Littenberg}},\ }\href {\doibase 10.1088/0264-9381/32/13/135012} {\bibfield
  {journal} {\bibinfo  {journal} {Class.\ Quant.\ Grav.}\ }\textbf {\bibinfo
  {volume} {32}},\ \bibinfo {pages} {135012} (\bibinfo {year} {2015})},\
  \Eprint {http://arxiv.org/abs/1410.3835} {arXiv:1410.3835 [gr-qc]}
  \BibitemShut {NoStop}%
\bibitem [{\citenamefont {Chatziioannou}\ \emph {et~al.}(2019)\citenamefont
  {Chatziioannou}, \citenamefont {Haster}, \citenamefont {Littenberg},
  \citenamefont {Farr}, \citenamefont {Ghonge}, \citenamefont {Millhouse},
  \citenamefont {Clark},\ and\ \citenamefont
  {Cornish}}]{Chatziioannou:2019zvs}%
  \BibitemOpen
  \bibfield  {author} {\bibinfo {author} {\bibfnamefont {K.}~\bibnamefont
  {Chatziioannou}}, \bibinfo {author} {\bibfnamefont {C.-J.}\ \bibnamefont
  {Haster}}, \bibinfo {author} {\bibfnamefont {T.~B.}\ \bibnamefont
  {Littenberg}}, \bibinfo {author} {\bibfnamefont {W.~M.}\ \bibnamefont
  {Farr}}, \bibinfo {author} {\bibfnamefont {S.}~\bibnamefont {Ghonge}},
  \bibinfo {author} {\bibfnamefont {M.}~\bibnamefont {Millhouse}}, \bibinfo
  {author} {\bibfnamefont {J.~A.}\ \bibnamefont {Clark}}, \ and\ \bibinfo
  {author} {\bibfnamefont {N.}~\bibnamefont {Cornish}},\ }\href {\doibase
  10.1103/PhysRevD.100.104004} {\bibfield  {journal} {\bibinfo  {journal}
  {Phys.\ Rev.\ D}\ }\textbf {\bibinfo {volume} {100}},\ \bibinfo {pages}
  {104004} (\bibinfo {year} {2019})},\ \Eprint
  {http://arxiv.org/abs/1907.06540} {arXiv:1907.06540 [gr-qc]} \BibitemShut
  {NoStop}%
\bibitem [{\citenamefont {{LIGO Scientific Collaboration and Virgo
  Collaboration}}(2019{\natexlab{a}})}]{GWTC1_PE_release}%
  \BibitemOpen
  \bibfield  {author} {\bibinfo {author} {\bibnamefont {{LIGO Scientific
  Collaboration and Virgo Collaboration}}},\ }\href {\doibase
  10.7935/KSX7-QQ51} {\enquote {\bibinfo {title} {{Parameter Estimation
  samples, Power Spectral Densities and Calibration Uncertainty Envelope
  release for GWTC-1}},}\ } (\bibinfo {year} {2019}{\natexlab{a}})\BibitemShut
  {NoStop}%
\bibitem [{\citenamefont {{LIGO Scientific Collaboration and Virgo
  Collaboration}}(2020{\natexlab{a}})}]{GW190425_PE_release}%
  \BibitemOpen
  \bibfield  {author} {\bibinfo {author} {\bibnamefont {{LIGO Scientific
  Collaboration and Virgo Collaboration}}},\ }\href
  {https://dcc.ligo.org/LIGO-P2000026/public} {\enquote {\bibinfo {title}
  {{Parameter estimation sample release for GW190425}},}\ } (\bibinfo {year}
  {2020}{\natexlab{a}})\BibitemShut {NoStop}%
\bibitem [{\citenamefont {{LIGO Scientific Collaboration and Virgo
  Collaboration}}(2020{\natexlab{b}})}]{GW190412_PE_release}%
  \BibitemOpen
  \bibfield  {author} {\bibinfo {author} {\bibnamefont {{LIGO Scientific
  Collaboration and Virgo Collaboration}}},\ }\href
  {https://dcc.ligo.org/LIGO-P190412/public} {\enquote {\bibinfo {title}
  {{Parameter estimation sample release for GW190412}},}\ } (\bibinfo {year}
  {2020}{\natexlab{b}})\BibitemShut {NoStop}%
\bibitem [{\citenamefont {Farr}\ \emph {et~al.}(2015)\citenamefont {Farr},
  \citenamefont {Farr},\ and\ \citenamefont
  {Littenberg}}]{SplineCalMarg-T1400682}%
  \BibitemOpen
  \bibfield  {author} {\bibinfo {author} {\bibfnamefont {W.~M.}\ \bibnamefont
  {Farr}}, \bibinfo {author} {\bibfnamefont {B.}~\bibnamefont {Farr}}, \ and\
  \bibinfo {author} {\bibfnamefont {T.}~\bibnamefont {Littenberg}},\ }\href
  {https://dcc.ligo.org/LIGO-P1500262/public} {\emph {\bibinfo {title}
  {Modelling Calibration Errors In CBC Waveforms}}},\ \bibinfo {type} {Tech.
  Rep.}\ \bibinfo {number}
  {\href{https://dcc.ligo.org/LIGO-P1500262/public}{LIGO}-T1400682}\ (\bibinfo
  {institution} {{LIGO} Project},\ \bibinfo {year} {2015})\BibitemShut
  {NoStop}%
\bibitem [{\citenamefont {Cahillane}\ \emph {et~al.}(2017)\citenamefont
  {Cahillane} \emph {et~al.}}]{Cahillane:2017vkb}%
  \BibitemOpen
  \bibfield  {author} {\bibinfo {author} {\bibfnamefont {C.}~\bibnamefont
  {Cahillane}} \emph {et~al.},\ }\href {\doibase 10.1103/PhysRevD.96.102001}
  {\bibfield  {journal} {\bibinfo  {journal} {Phys. Rev. D}\ }\textbf {\bibinfo
  {volume} {96}},\ \bibinfo {pages} {102001} (\bibinfo {year} {2017})},\
  \Eprint {http://arxiv.org/abs/1708.03023} {arXiv:1708.03023 [astro-ph.IM]}
  \BibitemShut {NoStop}%
\bibitem [{\citenamefont {{LIGO Scientific Collaboration}}(2018)}]{GWOSC_O1}%
  \BibitemOpen
  \bibfield  {author} {\bibinfo {author} {\bibnamefont {{LIGO Scientific
  Collaboration}}},\ }\href {\doibase 10.7935/K57P8W9D} {\enquote {\bibinfo
  {title} {{The O1 Data Release}},}\ } (\bibinfo {year} {2018})\BibitemShut
  {NoStop}%
\bibitem [{\citenamefont {{LIGO Scientific Collaboration and Virgo
  Collaboration}}(2019{\natexlab{b}})}]{GWOSC_O2}%
  \BibitemOpen
  \bibfield  {author} {\bibinfo {author} {\bibnamefont {{LIGO Scientific
  Collaboration and Virgo Collaboration}}},\ }\href {\doibase
  10.7935/CA75-FM95} {\enquote {\bibinfo {title} {{The O2 Data Release}},}\ }
  (\bibinfo {year} {2019}{\natexlab{b}})\BibitemShut {NoStop}%
\bibitem [{\citenamefont {{LIGO Scientific Collaboration and Virgo
  Collaboration}}(2020{\natexlab{c}})}]{GWOSC_GW190425}%
  \BibitemOpen
  \bibfield  {author} {\bibinfo {author} {\bibnamefont {{LIGO Scientific
  Collaboration and Virgo Collaboration}}},\ }\href {\doibase
  10.7935/ggb8-1v94} {\enquote {\bibinfo {title} {{The GW190425 Data
  Release}},}\ } (\bibinfo {year} {2020}{\natexlab{c}})\BibitemShut {NoStop}%
\bibitem [{\citenamefont {{LIGO Scientific Collaboration and Virgo
  Collaboration}}(2020{\natexlab{d}})}]{GWOSC_GW190412}%
  \BibitemOpen
  \bibfield  {author} {\bibinfo {author} {\bibnamefont {{LIGO Scientific
  Collaboration and Virgo Collaboration}}},\ }\href {\doibase
  10.7935/20yv-ka61} {\enquote {\bibinfo {title} {{The GW190412 Data
  Release}},}\ } (\bibinfo {year} {2020}{\natexlab{d}})\BibitemShut {NoStop}%
\bibitem [{\citenamefont {Vallisneri}\ \emph {et~al.}(2015)\citenamefont
  {Vallisneri}, \citenamefont {Kanner}, \citenamefont {Williams}, \citenamefont
  {Weinstein},\ and\ \citenamefont {Stephens}}]{Vallisneri:2014vxa}%
  \BibitemOpen
  \bibfield  {author} {\bibinfo {author} {\bibfnamefont {M.}~\bibnamefont
  {Vallisneri}}, \bibinfo {author} {\bibfnamefont {J.}~\bibnamefont {Kanner}},
  \bibinfo {author} {\bibfnamefont {R.}~\bibnamefont {Williams}}, \bibinfo
  {author} {\bibfnamefont {A.}~\bibnamefont {Weinstein}}, \ and\ \bibinfo
  {author} {\bibfnamefont {B.}~\bibnamefont {Stephens}},\ }\href {\doibase
  10.1088/1742-6596/610/1/012021} {\bibfield  {journal} {\bibinfo  {journal}
  {J.\ Phys.\ Conf.\ Ser.}\ }\textbf {\bibinfo {volume} {610}},\ \bibinfo
  {pages} {012021} (\bibinfo {year} {2015})},\ \Eprint
  {http://arxiv.org/abs/1410.4839} {arXiv:1410.4839 [gr-qc]} \BibitemShut
  {NoStop}%
\bibitem [{\citenamefont {Abbott}\ \emph
  {et~al.}(2019{\natexlab{d}})\citenamefont {Abbott} \emph
  {et~al.}}]{Abbott:2019ebz}%
  \BibitemOpen
  \bibfield  {author} {\bibinfo {author} {\bibfnamefont {R.}~\bibnamefont
  {Abbott}} \emph {et~al.} (\bibinfo {collaboration} {LIGO Scientific
  Collaboration, Virgo Collaboration}),\ }\href@noop {} {\  (\bibinfo {year}
  {2019}{\natexlab{d}})},\ \Eprint {http://arxiv.org/abs/1912.11716}
  {arXiv:1912.11716 [gr-qc]} \BibitemShut {NoStop}%
\bibitem [{\citenamefont {Abbott}\ \emph
  {et~al.}(2018{\natexlab{a}})\citenamefont {Abbott} \emph
  {et~al.}}]{Abbott:2018exr}%
  \BibitemOpen
  \bibfield  {author} {\bibinfo {author} {\bibfnamefont {B.}~\bibnamefont
  {Abbott}} \emph {et~al.} (\bibinfo {collaboration} {LIGO Scientific
  Collaboration, Virgo Collaboration}),\ }\href {\doibase
  10.1103/PhysRevLett.121.161101} {\bibfield  {journal} {\bibinfo  {journal}
  {Phys. Rev. Lett.}\ }\textbf {\bibinfo {volume} {121}},\ \bibinfo {pages}
  {161101} (\bibinfo {year} {2018}{\natexlab{a}})},\ \Eprint
  {http://arxiv.org/abs/1805.11581} {arXiv:1805.11581 [gr-qc]} \BibitemShut
  {NoStop}%
\bibitem [{\citenamefont {Abbott}\ \emph
  {et~al.}(2020{\natexlab{a}})\citenamefont {Abbott} \emph
  {et~al.}}]{Abbott:2020uma}%
  \BibitemOpen
  \bibfield  {author} {\bibinfo {author} {\bibfnamefont {B.}~\bibnamefont
  {Abbott}} \emph {et~al.} (\bibinfo {collaboration} {LIGO Scientific
  Collaboration, Virgo Collaboration}),\ }\href {\doibase
  10.3847/2041-8213/ab75f5} {\bibfield  {journal} {\bibinfo  {journal}
  {Astrophys.\ J.\ Lett.}\ }\textbf {\bibinfo {volume} {892}},\ \bibinfo
  {pages} {L3} (\bibinfo {year} {2020}{\natexlab{a}})},\ \Eprint
  {http://arxiv.org/abs/2001.01761} {arXiv:2001.01761 [astro-ph.HE]}
  \BibitemShut {NoStop}%
\bibitem [{\citenamefont {Abbott}\ \emph
  {et~al.}(2020{\natexlab{b}})\citenamefont {Abbott} \emph
  {et~al.}}]{LIGOScientific:2020stg}%
  \BibitemOpen
  \bibfield  {author} {\bibinfo {author} {\bibfnamefont {R.}~\bibnamefont
  {Abbott}} \emph {et~al.} (\bibinfo {collaboration} {LIGO Scientific
  Collaboration, Virgo Collaboration}),\ }\href@noop {} {\  (\bibinfo {year}
  {2020}{\natexlab{b}})},\ \Eprint {http://arxiv.org/abs/2004.08342}
  {arXiv:2004.08342 [astro-ph.HE]} \BibitemShut {NoStop}%
\bibitem [{\citenamefont {Chatziioannou}\ \emph {et~al.}(2018)\citenamefont
  {Chatziioannou}, \citenamefont {Haster},\ and\ \citenamefont
  {Zimmerman}}]{Chatziioannou:2018vzf}%
  \BibitemOpen
  \bibfield  {author} {\bibinfo {author} {\bibfnamefont {K.}~\bibnamefont
  {Chatziioannou}}, \bibinfo {author} {\bibfnamefont {C.-J.}\ \bibnamefont
  {Haster}}, \ and\ \bibinfo {author} {\bibfnamefont {A.}~\bibnamefont
  {Zimmerman}},\ }\href {\doibase 10.1103/PhysRevD.97.104036} {\bibfield
  {journal} {\bibinfo  {journal} {Phys.\ Rev.\ D}\ }\textbf {\bibinfo {volume}
  {97}},\ \bibinfo {pages} {104036} (\bibinfo {year} {2018})},\ \Eprint
  {http://arxiv.org/abs/1804.03221} {arXiv:1804.03221 [gr-qc]} \BibitemShut
  {NoStop}%
\bibitem [{\citenamefont {Dickey}(1971)}]{dickey1971}%
  \BibitemOpen
  \bibfield  {author} {\bibinfo {author} {\bibfnamefont {J.~M.}\ \bibnamefont
  {Dickey}},\ }\href {\doibase 10.1214/aoms/1177693507} {\bibfield  {journal}
  {\bibinfo  {journal} {Ann. Math. Statist.}\ }\textbf {\bibinfo {volume}
  {42}},\ \bibinfo {pages} {204} (\bibinfo {year} {1971})}\BibitemShut
  {NoStop}%
\bibitem [{\citenamefont {Verdinelli}\ and\ \citenamefont
  {Wasserman}(1995)}]{BF_SavageDickey}%
  \BibitemOpen
  \bibfield  {author} {\bibinfo {author} {\bibfnamefont {I.}~\bibnamefont
  {Verdinelli}}\ and\ \bibinfo {author} {\bibfnamefont {L.}~\bibnamefont
  {Wasserman}},\ }\href {\doibase 10.1080/01621459.1995.10476554} {\bibfield
  {journal} {\bibinfo  {journal} {Journal of the American Statistical
  Association}\ }\textbf {\bibinfo {volume} {90}},\ \bibinfo {pages} {614}
  (\bibinfo {year} {1995})}\BibitemShut {NoStop}%
\bibitem [{\citenamefont {Abbott}\ \emph
  {et~al.}(2018{\natexlab{b}})\citenamefont {Abbott} \emph
  {et~al.}}]{Aasi:2013wya}%
  \BibitemOpen
  \bibfield  {author} {\bibinfo {author} {\bibfnamefont {B.}~\bibnamefont
  {Abbott}} \emph {et~al.} (\bibinfo {collaboration} {KAGRA Collaboration,LIGO
  Scientific Collaboration, Virgo Collaboration}),\ }\href {\doibase
  10.1007/s41114-018-0012-9} {\bibfield  {journal} {\bibinfo  {journal} {Living
  Rev. Rel.}\ }\textbf {\bibinfo {volume} {21}},\ \bibinfo {pages} {3}
  (\bibinfo {year} {2018}{\natexlab{b}})},\ \Eprint
  {http://arxiv.org/abs/1304.0670} {arXiv:1304.0670 [gr-qc]} \BibitemShut
  {NoStop}%
\bibitem [{\citenamefont {Zackay}\ \emph
  {et~al.}(2019{\natexlab{a}})\citenamefont {Zackay}, \citenamefont
  {Venumadhav}, \citenamefont {Dai}, \citenamefont {Roulet},\ and\
  \citenamefont {Zaldarriaga}}]{Zackay:2019tzo}%
  \BibitemOpen
  \bibfield  {author} {\bibinfo {author} {\bibfnamefont {B.}~\bibnamefont
  {Zackay}}, \bibinfo {author} {\bibfnamefont {T.}~\bibnamefont {Venumadhav}},
  \bibinfo {author} {\bibfnamefont {L.}~\bibnamefont {Dai}}, \bibinfo {author}
  {\bibfnamefont {J.}~\bibnamefont {Roulet}}, \ and\ \bibinfo {author}
  {\bibfnamefont {M.}~\bibnamefont {Zaldarriaga}},\ }\href {\doibase
  10.1103/PhysRevD.100.023007} {\bibfield  {journal} {\bibinfo  {journal}
  {Phys.\ Rev.\ D}\ }\textbf {\bibinfo {volume} {100}},\ \bibinfo {pages}
  {023007} (\bibinfo {year} {2019}{\natexlab{a}})},\ \Eprint
  {http://arxiv.org/abs/1902.10331} {arXiv:1902.10331 [astro-ph.HE]}
  \BibitemShut {NoStop}%
\bibitem [{\citenamefont {Venumadhav}\ \emph {et~al.}(2020)\citenamefont
  {Venumadhav}, \citenamefont {Zackay}, \citenamefont {Roulet}, \citenamefont
  {Dai},\ and\ \citenamefont {Zaldarriaga}}]{Venumadhav:2019lyq}%
  \BibitemOpen
  \bibfield  {author} {\bibinfo {author} {\bibfnamefont {T.}~\bibnamefont
  {Venumadhav}}, \bibinfo {author} {\bibfnamefont {B.}~\bibnamefont {Zackay}},
  \bibinfo {author} {\bibfnamefont {J.}~\bibnamefont {Roulet}}, \bibinfo
  {author} {\bibfnamefont {L.}~\bibnamefont {Dai}}, \ and\ \bibinfo {author}
  {\bibfnamefont {M.}~\bibnamefont {Zaldarriaga}},\ }\href {\doibase
  10.1103/PhysRevD.101.083030} {\bibfield  {journal} {\bibinfo  {journal}
  {Phys. Rev. D}\ }\textbf {\bibinfo {volume} {101}},\ \bibinfo {pages}
  {083030} (\bibinfo {year} {2020})},\ \Eprint
  {http://arxiv.org/abs/1904.07214} {arXiv:1904.07214 [astro-ph.HE]}
  \BibitemShut {NoStop}%
\bibitem [{\citenamefont {Nitz}\ \emph {et~al.}(2019)\citenamefont {Nitz},
  \citenamefont {Dent}, \citenamefont {Davies}, \citenamefont {Kumar},
  \citenamefont {Capano}, \citenamefont {Harry}, \citenamefont {Mozzon},
  \citenamefont {Nuttall}, \citenamefont {Lundgren},\ and\ \citenamefont
  {T{\'a}pai}}]{Nitz:2019hdf}%
  \BibitemOpen
  \bibfield  {author} {\bibinfo {author} {\bibfnamefont {A.~H.}\ \bibnamefont
  {Nitz}}, \bibinfo {author} {\bibfnamefont {T.}~\bibnamefont {Dent}}, \bibinfo
  {author} {\bibfnamefont {G.~S.}\ \bibnamefont {Davies}}, \bibinfo {author}
  {\bibfnamefont {S.}~\bibnamefont {Kumar}}, \bibinfo {author} {\bibfnamefont
  {C.~D.}\ \bibnamefont {Capano}}, \bibinfo {author} {\bibfnamefont
  {I.}~\bibnamefont {Harry}}, \bibinfo {author} {\bibfnamefont
  {S.}~\bibnamefont {Mozzon}}, \bibinfo {author} {\bibfnamefont
  {L.}~\bibnamefont {Nuttall}}, \bibinfo {author} {\bibfnamefont
  {A.}~\bibnamefont {Lundgren}}, \ and\ \bibinfo {author} {\bibfnamefont
  {M.}~\bibnamefont {T{\'a}pai}},\ }\href {\doibase 10.3847/1538-4357/ab733f}
  {\bibfield  {journal} {\bibinfo  {journal} {Astrophys.\ J.}\ }\textbf
  {\bibinfo {volume} {891}},\ \bibinfo {pages} {123} (\bibinfo {year}
  {2019})},\ \Eprint {http://arxiv.org/abs/1910.05331} {arXiv:1910.05331
  [astro-ph.HE]} \BibitemShut {NoStop}%
\bibitem [{\citenamefont {Zackay}\ \emph
  {et~al.}(2019{\natexlab{b}})\citenamefont {Zackay}, \citenamefont {Dai},
  \citenamefont {Venumadhav}, \citenamefont {Roulet},\ and\ \citenamefont
  {Zaldarriaga}}]{Zackay:2019btq}%
  \BibitemOpen
  \bibfield  {author} {\bibinfo {author} {\bibfnamefont {B.}~\bibnamefont
  {Zackay}}, \bibinfo {author} {\bibfnamefont {L.}~\bibnamefont {Dai}},
  \bibinfo {author} {\bibfnamefont {T.}~\bibnamefont {Venumadhav}}, \bibinfo
  {author} {\bibfnamefont {J.}~\bibnamefont {Roulet}}, \ and\ \bibinfo {author}
  {\bibfnamefont {M.}~\bibnamefont {Zaldarriaga}},\ }\href@noop {} {\
  (\bibinfo {year} {2019}{\natexlab{b}})},\ \Eprint
  {http://arxiv.org/abs/1910.09528} {arXiv:1910.09528 [astro-ph.HE]}
  \BibitemShut {NoStop}%
\bibitem [{\citenamefont {Galaudage}\ \emph {et~al.}(2019)\citenamefont
  {Galaudage}, \citenamefont {Talbot},\ and\ \citenamefont
  {Thrane}}]{Galaudage:2019jdx}%
  \BibitemOpen
  \bibfield  {author} {\bibinfo {author} {\bibfnamefont {S.}~\bibnamefont
  {Galaudage}}, \bibinfo {author} {\bibfnamefont {C.}~\bibnamefont {Talbot}}, \
  and\ \bibinfo {author} {\bibfnamefont {E.}~\bibnamefont {Thrane}},\
  }\href@noop {} {\  (\bibinfo {year} {2019})},\ \Eprint
  {http://arxiv.org/abs/1912.09708} {arXiv:1912.09708 [astro-ph.HE]}
  \BibitemShut {NoStop}%
\bibitem [{\citenamefont {Huang}\ \emph {et~al.}(2020)\citenamefont {Huang},
  \citenamefont {Haster}, \citenamefont {Vitale}, \citenamefont {Zimmerman},
  \citenamefont {Roulet}, \citenamefont {Venumadhav}, \citenamefont {Zackay},
  \citenamefont {Dai},\ and\ \citenamefont {Zaldarriaga}}]{Huang:2020ysn}%
  \BibitemOpen
  \bibfield  {author} {\bibinfo {author} {\bibfnamefont {Y.}~\bibnamefont
  {Huang}}, \bibinfo {author} {\bibfnamefont {C.-J.}\ \bibnamefont {Haster}},
  \bibinfo {author} {\bibfnamefont {S.}~\bibnamefont {Vitale}}, \bibinfo
  {author} {\bibfnamefont {A.}~\bibnamefont {Zimmerman}}, \bibinfo {author}
  {\bibfnamefont {J.}~\bibnamefont {Roulet}}, \bibinfo {author} {\bibfnamefont
  {T.}~\bibnamefont {Venumadhav}}, \bibinfo {author} {\bibfnamefont
  {B.}~\bibnamefont {Zackay}}, \bibinfo {author} {\bibfnamefont
  {L.}~\bibnamefont {Dai}}, \ and\ \bibinfo {author} {\bibfnamefont
  {M.}~\bibnamefont {Zaldarriaga}},\ }\href@noop {} {\  (\bibinfo {year}
  {2020})},\ \Eprint {http://arxiv.org/abs/2003.04513} {arXiv:2003.04513
  [gr-qc]} \BibitemShut {NoStop}%
\bibitem [{Min()}]{MinusTwoPN}%
  \BibitemOpen
  \href@noop {} {}\bibinfo {note} {{The single-\ac{PN}-order analyses in
  ~\cite{Abbott:2018lct,LIGOScientific:2019fpa,LIGOScientific:2020stg} also
  cover $\varphi_{-2}$ probing the presence of leading-order dipolar \acp{GW},
  a term not included in the current formalism of the variable $\pi$ analysis.
  The fractional widths of these constraints are
  $|\varphi_{-2}|<2\times10^{-5}$~\cite{Abbott:2018lct},
  $|\varphi_{-2}|<3\times10^{-3}$~\cite{LIGOScientific:2019fpa} and
  $|\varphi_{-2}|<2\times10^{-2}$~\cite{LIGOScientific:2020stg}
  respectively.}}\BibitemShut {Stop}%
\bibitem [{\citenamefont {{LIGO Scientific Collaboration and Virgo
  Collaboration}}(2019{\natexlab{c}})}]{GWTC-1_TGR_dataRelease}%
  \BibitemOpen
  \bibfield  {author} {\bibinfo {author} {\bibnamefont {{LIGO Scientific
  Collaboration and Virgo Collaboration}}},\ }\href
  {https://dcc.ligo.org/LIGO-P1900087/public} {\enquote {\bibinfo {title}
  {{Data release for testing GR with GWTC-1 events}},}\ } (\bibinfo {year}
  {2019}{\natexlab{c}})\BibitemShut {NoStop}%
\bibitem [{\citenamefont {Nitz}\ \emph {et~al.}(2017)\citenamefont {Nitz},
  \citenamefont {Dent}, \citenamefont {Dal~Canton}, \citenamefont {Fairhurst},\
  and\ \citenamefont {Brown}}]{Nitz:2017svb}%
  \BibitemOpen
  \bibfield  {author} {\bibinfo {author} {\bibfnamefont {A.~H.}\ \bibnamefont
  {Nitz}}, \bibinfo {author} {\bibfnamefont {T.}~\bibnamefont {Dent}}, \bibinfo
  {author} {\bibfnamefont {T.}~\bibnamefont {Dal~Canton}}, \bibinfo {author}
  {\bibfnamefont {S.}~\bibnamefont {Fairhurst}}, \ and\ \bibinfo {author}
  {\bibfnamefont {D.~A.}\ \bibnamefont {Brown}},\ }\href {\doibase
  10.3847/1538-4357/aa8f50} {\bibfield  {journal} {\bibinfo  {journal}
  {Astrophys. J.}\ }\textbf {\bibinfo {volume} {849}},\ \bibinfo {pages} {118}
  (\bibinfo {year} {2017})},\ \Eprint {http://arxiv.org/abs/1705.01513}
  {arXiv:1705.01513 [gr-qc]} \BibitemShut {NoStop}%
\bibitem [{\citenamefont {Usman}\ \emph {et~al.}(2016)\citenamefont {Usman}
  \emph {et~al.}}]{Usman:2015kfa}%
  \BibitemOpen
  \bibfield  {author} {\bibinfo {author} {\bibfnamefont {S.~A.}\ \bibnamefont
  {Usman}} \emph {et~al.},\ }\href {\doibase 10.1088/0264-9381/33/21/215004}
  {\bibfield  {journal} {\bibinfo  {journal} {Class. Quant. Grav.}\ }\textbf
  {\bibinfo {volume} {33}},\ \bibinfo {pages} {215004} (\bibinfo {year}
  {2016})},\ \Eprint {http://arxiv.org/abs/1508.02357} {arXiv:1508.02357
  [gr-qc]} \BibitemShut {NoStop}%
\bibitem [{\citenamefont {Dal~Canton}\ \emph {et~al.}(2014)\citenamefont
  {Dal~Canton} \emph {et~al.}}]{Canton:2014ena}%
  \BibitemOpen
  \bibfield  {author} {\bibinfo {author} {\bibfnamefont {T.}~\bibnamefont
  {Dal~Canton}} \emph {et~al.},\ }\href {\doibase 10.1103/PhysRevD.90.082004}
  {\bibfield  {journal} {\bibinfo  {journal} {Phys. Rev. D}\ }\textbf {\bibinfo
  {volume} {90}},\ \bibinfo {pages} {082004} (\bibinfo {year} {2014})},\
  \Eprint {http://arxiv.org/abs/1405.6731} {arXiv:1405.6731 [gr-qc]}
  \BibitemShut {NoStop}%
\bibitem [{\citenamefont {Nitz}\ \emph {et~al.}(2020)\citenamefont {Nitz} \emph
  {et~al.}}]{PyCBC_zenodo}%
  \BibitemOpen
  \bibfield  {author} {\bibinfo {author} {\bibfnamefont {A.~H.}\ \bibnamefont
  {Nitz}} \emph {et~al.},\ }\href {\doibase 10.5281/zenodo.596388} {\enquote
  {\bibinfo {title} {gwastro/pycbc},}\ } (\bibinfo {year} {2020})\BibitemShut
  {NoStop}%
\bibitem [{\citenamefont {Isi}\ \emph {et~al.}(2018)\citenamefont {Isi},
  \citenamefont {Smith}, \citenamefont {Vitale}, \citenamefont {Massinger},
  \citenamefont {Kanner},\ and\ \citenamefont {Vajpeyi}}]{Isi:2018vst}%
  \BibitemOpen
  \bibfield  {author} {\bibinfo {author} {\bibfnamefont {M.}~\bibnamefont
  {Isi}}, \bibinfo {author} {\bibfnamefont {R.}~\bibnamefont {Smith}}, \bibinfo
  {author} {\bibfnamefont {S.}~\bibnamefont {Vitale}}, \bibinfo {author}
  {\bibfnamefont {T.}~\bibnamefont {Massinger}}, \bibinfo {author}
  {\bibfnamefont {J.}~\bibnamefont {Kanner}}, \ and\ \bibinfo {author}
  {\bibfnamefont {A.}~\bibnamefont {Vajpeyi}},\ }\href {\doibase
  10.1103/PhysRevD.98.042007} {\bibfield  {journal} {\bibinfo  {journal} {Phys.
  Rev. D}\ }\textbf {\bibinfo {volume} {98}},\ \bibinfo {pages} {042007}
  (\bibinfo {year} {2018})},\ \Eprint {http://arxiv.org/abs/1803.09783}
  {arXiv:1803.09783 [gr-qc]} \BibitemShut {NoStop}%
\bibitem [{\citenamefont {Will}(1998)}]{Will:1997bb}%
  \BibitemOpen
  \bibfield  {author} {\bibinfo {author} {\bibfnamefont {C.~M.}\ \bibnamefont
  {Will}},\ }\href {\doibase 10.1103/PhysRevD.57.2061} {\bibfield  {journal}
  {\bibinfo  {journal} {Phys. Rev. D}\ }\textbf {\bibinfo {volume} {57}},\
  \bibinfo {pages} {2061} (\bibinfo {year} {1998})},\ \Eprint
  {http://arxiv.org/abs/gr-qc/9709011} {arXiv:gr-qc/9709011} \BibitemShut
  {NoStop}%
\bibitem [{\citenamefont {Abbott}\ \emph
  {et~al.}(2019{\natexlab{e}})\citenamefont {Abbott} \emph
  {et~al.}}]{Pisarski:2019vxw}%
  \BibitemOpen
  \bibfield  {author} {\bibinfo {author} {\bibfnamefont {B.}~\bibnamefont
  {Abbott}} \emph {et~al.} (\bibinfo {collaboration} {LIGO Scientific
  Collaboration, Virgo Collaboration}),\ }\href {\doibase
  10.1103/PhysRevD.100.024004} {\bibfield  {journal} {\bibinfo  {journal}
  {Phys. Rev. D}\ }\textbf {\bibinfo {volume} {100}},\ \bibinfo {pages}
  {024004} (\bibinfo {year} {2019}{\natexlab{e}})},\ \Eprint
  {http://arxiv.org/abs/1903.01901} {arXiv:1903.01901 [astro-ph.HE]}
  \BibitemShut {NoStop}%
\bibitem [{\citenamefont {Abbott}\ \emph
  {et~al.}(2019{\natexlab{f}})\citenamefont {Abbott} \emph
  {et~al.}}]{Abbott:2019bed}%
  \BibitemOpen
  \bibfield  {author} {\bibinfo {author} {\bibfnamefont {B.}~\bibnamefont
  {Abbott}} \emph {et~al.} (\bibinfo {collaboration} {LIGO Scientific
  Collaboration, Virgo Collaboration}),\ }\href {\doibase
  10.1103/PhysRevD.99.122002} {\bibfield  {journal} {\bibinfo  {journal} {Phys.
  Rev. D}\ }\textbf {\bibinfo {volume} {99}},\ \bibinfo {pages} {122002}
  (\bibinfo {year} {2019}{\natexlab{f}})},\ \Eprint
  {http://arxiv.org/abs/1902.08442} {arXiv:1902.08442 [gr-qc]} \BibitemShut
  {NoStop}%
\bibitem [{\citenamefont {Abbott}\ \emph
  {et~al.}(2019{\natexlab{g}})\citenamefont {Abbott} \emph
  {et~al.}}]{Authors:2019ztc}%
  \BibitemOpen
  \bibfield  {author} {\bibinfo {author} {\bibfnamefont {B.}~\bibnamefont
  {Abbott}} \emph {et~al.} (\bibinfo {collaboration} {LIGO Scientific
  Collaboration, Virgo Collaboration}),\ }\href {\doibase
  10.3847/1538-4357/ab20cb} {\bibfield  {journal} {\bibinfo  {journal}
  {Astrophys. J.}\ }\textbf {\bibinfo {volume} {879}},\ \bibinfo {pages} {10}
  (\bibinfo {year} {2019}{\natexlab{g}})},\ \Eprint
  {http://arxiv.org/abs/1902.08507} {arXiv:1902.08507 [astro-ph.HE]}
  \BibitemShut {NoStop}%
\bibitem [{\citenamefont {Sesana}(2016)}]{Sesana:2016ljz}%
  \BibitemOpen
  \bibfield  {author} {\bibinfo {author} {\bibfnamefont {A.}~\bibnamefont
  {Sesana}},\ }\href {\doibase 10.1103/PhysRevLett.116.231102} {\bibfield
  {journal} {\bibinfo  {journal} {Phys. Rev. Lett.}\ }\textbf {\bibinfo
  {volume} {116}},\ \bibinfo {pages} {231102} (\bibinfo {year} {2016})},\
  \Eprint {http://arxiv.org/abs/1602.06951} {arXiv:1602.06951 [gr-qc]}
  \BibitemShut {NoStop}%
\bibitem [{\citenamefont {Vitale}(2016)}]{Vitale:2016rfr}%
  \BibitemOpen
  \bibfield  {author} {\bibinfo {author} {\bibfnamefont {S.}~\bibnamefont
  {Vitale}},\ }\href {\doibase 10.1103/PhysRevLett.117.051102} {\bibfield
  {journal} {\bibinfo  {journal} {Phys. Rev. Lett.}\ }\textbf {\bibinfo
  {volume} {117}},\ \bibinfo {pages} {051102} (\bibinfo {year} {2016})},\
  \Eprint {http://arxiv.org/abs/1605.01037} {arXiv:1605.01037 [gr-qc]}
  \BibitemShut {NoStop}%
\bibitem [{\citenamefont {Chamberlain}\ and\ \citenamefont
  {Yunes}(2017)}]{Chamberlain:2017fjl}%
  \BibitemOpen
  \bibfield  {author} {\bibinfo {author} {\bibfnamefont {K.}~\bibnamefont
  {Chamberlain}}\ and\ \bibinfo {author} {\bibfnamefont {N.}~\bibnamefont
  {Yunes}},\ }\href {\doibase 10.1103/PhysRevD.96.084039} {\bibfield  {journal}
  {\bibinfo  {journal} {Phys. Rev. D}\ }\textbf {\bibinfo {volume} {96}},\
  \bibinfo {pages} {084039} (\bibinfo {year} {2017})},\ \Eprint
  {http://arxiv.org/abs/1704.08268} {arXiv:1704.08268 [gr-qc]} \BibitemShut
  {NoStop}%
\bibitem [{\citenamefont {Carson}\ and\ \citenamefont
  {Yagi}(2020{\natexlab{a}})}]{Carson:2019kkh}%
  \BibitemOpen
  \bibfield  {author} {\bibinfo {author} {\bibfnamefont {Z.}~\bibnamefont
  {Carson}}\ and\ \bibinfo {author} {\bibfnamefont {K.}~\bibnamefont {Yagi}},\
  }\href {\doibase 10.1103/PhysRevD.101.044047} {\bibfield  {journal} {\bibinfo
   {journal} {Phys. Rev. D}\ }\textbf {\bibinfo {volume} {101}},\ \bibinfo
  {pages} {044047} (\bibinfo {year} {2020}{\natexlab{a}})},\ \Eprint
  {http://arxiv.org/abs/1911.05258} {arXiv:1911.05258 [gr-qc]} \BibitemShut
  {NoStop}%
\bibitem [{\citenamefont {Carson}\ and\ \citenamefont
  {Yagi}(2020{\natexlab{b}})}]{Carson:2019rda}%
  \BibitemOpen
  \bibfield  {author} {\bibinfo {author} {\bibfnamefont {Z.}~\bibnamefont
  {Carson}}\ and\ \bibinfo {author} {\bibfnamefont {K.}~\bibnamefont {Yagi}},\
  }\href {\doibase 10.1088/1361-6382/ab5c9a} {\bibfield  {journal} {\bibinfo
  {journal} {Class. Quant. Grav.}\ }\textbf {\bibinfo {volume} {37}},\ \bibinfo
  {pages} {02LT01} (\bibinfo {year} {2020}{\natexlab{b}})},\ \Eprint
  {http://arxiv.org/abs/1905.13155} {arXiv:1905.13155 [gr-qc]} \BibitemShut
  {NoStop}%
\bibitem [{\citenamefont {Marsat}\ \emph {et~al.}(2020)\citenamefont {Marsat},
  \citenamefont {Baker},\ and\ \citenamefont {Dal~Canton}}]{Marsat:2020rtl}%
  \BibitemOpen
  \bibfield  {author} {\bibinfo {author} {\bibfnamefont {S.}~\bibnamefont
  {Marsat}}, \bibinfo {author} {\bibfnamefont {J.~G.}\ \bibnamefont {Baker}}, \
  and\ \bibinfo {author} {\bibfnamefont {T.}~\bibnamefont {Dal~Canton}},\
  }\href@noop {} {\  (\bibinfo {year} {2020})},\ \Eprint
  {http://arxiv.org/abs/2003.00357} {arXiv:2003.00357 [gr-qc]} \BibitemShut
  {NoStop}%
\bibitem [{\citenamefont {Toubiana}\ \emph {et~al.}(2020)\citenamefont
  {Toubiana}, \citenamefont {Marsat}, \citenamefont {Barausse}, \citenamefont
  {Babak},\ and\ \citenamefont {Baker}}]{Toubiana:2020vtf}%
  \BibitemOpen
  \bibfield  {author} {\bibinfo {author} {\bibfnamefont {A.}~\bibnamefont
  {Toubiana}}, \bibinfo {author} {\bibfnamefont {S.}~\bibnamefont {Marsat}},
  \bibinfo {author} {\bibfnamefont {E.}~\bibnamefont {Barausse}}, \bibinfo
  {author} {\bibfnamefont {S.}~\bibnamefont {Babak}}, \ and\ \bibinfo {author}
  {\bibfnamefont {J.}~\bibnamefont {Baker}},\ }\href@noop {} {\  (\bibinfo
  {year} {2020})},\ \Eprint {http://arxiv.org/abs/2004.03626} {arXiv:2004.03626
  [gr-qc]} \BibitemShut {NoStop}%
\bibitem [{\citenamefont {Kelley}\ \emph {et~al.}(2018)\citenamefont {Kelley},
  \citenamefont {Blecha}, \citenamefont {Hernquist}, \citenamefont {Sesana},\
  and\ \citenamefont {Taylor}}]{Kelley:2017vox}%
  \BibitemOpen
  \bibfield  {author} {\bibinfo {author} {\bibfnamefont {L.~Z.}\ \bibnamefont
  {Kelley}}, \bibinfo {author} {\bibfnamefont {L.}~\bibnamefont {Blecha}},
  \bibinfo {author} {\bibfnamefont {L.}~\bibnamefont {Hernquist}}, \bibinfo
  {author} {\bibfnamefont {A.}~\bibnamefont {Sesana}}, \ and\ \bibinfo {author}
  {\bibfnamefont {S.~R.}\ \bibnamefont {Taylor}},\ }\href {\doibase
  10.1093/mnras/sty689} {\bibfield  {journal} {\bibinfo  {journal} {Mon. Not.
  Roy. Astron. Soc.}\ }\textbf {\bibinfo {volume} {477}},\ \bibinfo {pages}
  {964} (\bibinfo {year} {2018})},\ \Eprint {http://arxiv.org/abs/1711.00075}
  {arXiv:1711.00075 [astro-ph.HE]} \BibitemShut {NoStop}%
\bibitem [{\citenamefont {Burke-Spolaor}\ \emph {et~al.}(2019)\citenamefont
  {Burke-Spolaor} \emph {et~al.}}]{Burke-Spolaor:2018bvk}%
  \BibitemOpen
  \bibfield  {author} {\bibinfo {author} {\bibfnamefont {S.}~\bibnamefont
  {Burke-Spolaor}} \emph {et~al.},\ }\href {\doibase 10.1007/s00159-019-0115-7}
  {\bibfield  {journal} {\bibinfo  {journal} {Astron. Astrophys. Rev.}\
  }\textbf {\bibinfo {volume} {27}},\ \bibinfo {pages} {5} (\bibinfo {year}
  {2019})},\ \Eprint {http://arxiv.org/abs/1811.08826} {arXiv:1811.08826
  [astro-ph.HE]} \BibitemShut {NoStop}%
\bibitem [{\citenamefont {Aggarwal}\ \emph {et~al.}(2019)\citenamefont
  {Aggarwal} \emph {et~al.}}]{Aggarwal:2018mgp}%
  \BibitemOpen
  \bibfield  {author} {\bibinfo {author} {\bibfnamefont {K.}~\bibnamefont
  {Aggarwal}} \emph {et~al.},\ }\href {\doibase 10.3847/1538-4357/ab2236}
  {\bibfield  {journal} {\bibinfo  {journal} {Astrophys. J.}\ }\textbf
  {\bibinfo {volume} {880}},\ \bibinfo {pages} {2} (\bibinfo {year} {2019})},\
  \Eprint {http://arxiv.org/abs/1812.11585} {arXiv:1812.11585 [astro-ph.GA]}
  \BibitemShut {NoStop}%
\bibitem [{\citenamefont {B{\'e}csy}\ and\ \citenamefont
  {Cornish}(2019)}]{Becsy:2019dim}%
  \BibitemOpen
  \bibfield  {author} {\bibinfo {author} {\bibfnamefont {B.}~\bibnamefont
  {B{\'e}csy}}\ and\ \bibinfo {author} {\bibfnamefont {N.~J.}\ \bibnamefont
  {Cornish}},\ }\href@noop {} {\  (\bibinfo {year} {2019})},\ \Eprint
  {http://arxiv.org/abs/1912.08807} {arXiv:1912.08807 [gr-qc]} \BibitemShut
  {NoStop}%
\bibitem [{\citenamefont {Gnocchi}\ \emph {et~al.}(2019)\citenamefont
  {Gnocchi}, \citenamefont {Maselli}, \citenamefont {Abdelsalhin},
  \citenamefont {Giacobbo},\ and\ \citenamefont {Mapelli}}]{Gnocchi:2019jzp}%
  \BibitemOpen
  \bibfield  {author} {\bibinfo {author} {\bibfnamefont {G.}~\bibnamefont
  {Gnocchi}}, \bibinfo {author} {\bibfnamefont {A.}~\bibnamefont {Maselli}},
  \bibinfo {author} {\bibfnamefont {T.}~\bibnamefont {Abdelsalhin}}, \bibinfo
  {author} {\bibfnamefont {N.}~\bibnamefont {Giacobbo}}, \ and\ \bibinfo
  {author} {\bibfnamefont {M.}~\bibnamefont {Mapelli}},\ }\href {\doibase
  10.1103/PhysRevD.100.064024} {\bibfield  {journal} {\bibinfo  {journal}
  {Phys. Rev. D}\ }\textbf {\bibinfo {volume} {100}},\ \bibinfo {pages}
  {064024} (\bibinfo {year} {2019})},\ \Eprint
  {http://arxiv.org/abs/1905.13460} {arXiv:1905.13460 [gr-qc]} \BibitemShut
  {NoStop}%
\bibitem [{\citenamefont {Oliphant}(06  )}]{numpy}%
  \BibitemOpen
  \bibfield  {author} {\bibinfo {author} {\bibfnamefont {T.}~\bibnamefont
  {Oliphant}},\ }\href {http://www.numpy.org/} {\enquote {\bibinfo {title}
  {{NumPy}: A guide to {NumPy}},}\ }\bibinfo {howpublished} {USA: Trelgol
  Publishing} (\bibinfo {year} {2006--})\BibitemShut {NoStop}%
\bibitem [{\citenamefont {Virtanen}\ \emph {et~al.}(2020)\citenamefont
  {Virtanen} \emph {et~al.}}]{Virtanen:2019joe}%
  \BibitemOpen
  \bibfield  {author} {\bibinfo {author} {\bibfnamefont {P.}~\bibnamefont
  {Virtanen}} \emph {et~al.},\ }\href {\doibase 10.1038/s41592-019-0686-2}
  {\bibfield  {journal} {\bibinfo  {journal} {Nature Meth.}\ }\textbf {\bibinfo
  {volume} {17}},\ \bibinfo {pages} {261} (\bibinfo {year} {2020})},\ \Eprint
  {http://arxiv.org/abs/1907.10121} {arXiv:1907.10121 [cs.MS]} \BibitemShut
  {NoStop}%
\bibitem [{\citenamefont {Hunter}(2007)}]{Hunter:2007ouj}%
  \BibitemOpen
  \bibfield  {author} {\bibinfo {author} {\bibfnamefont {J.~D.}\ \bibnamefont
  {Hunter}},\ }\href {\doibase 10.1109/MCSE.2007.55} {\bibfield  {journal}
  {\bibinfo  {journal} {Comput. Sci. Eng.}\ }\textbf {\bibinfo {volume} {9}},\
  \bibinfo {pages} {90} (\bibinfo {year} {2007})}\BibitemShut {NoStop}%
\end{thebibliography}%

\end{document}